%% file: AARv_NUMAGS_arXiv.tex
\RequirePackage{fix-cm}
\documentclass[natbib]{svjour3}                     % onecolumn (standard format)
\smartqed  % flush right qed marks, e.g. at end of proof
\usepackage{graphicx}
\usepackage{color, colortbl}
\usepackage[citecolor=blue,urlcolor=blue,breaklinks]{hyperref}
\definecolor{Gray}{gray}{0.9}
\definecolor{DarkRed}{rgb}{0.65,0.16,0.16}
\journalname{Astron Astrophys Rev}
\begin{document}

\title{Small solar system bodies as granular media\thanks{Made possible by the International Space Science Institute (ISSI, Bern) support to the international team ``Asteroids \& Self Gravitating Bodies as Granular Systems''
%Grants or other notes about the article that should go on the front page should be placed here. General acknowledgments should be placed at the end of the article.
}}


\author{D. Hestroffer         
\and
	P. S{\'a}nchez
\and
	L. Staron
\and
	A. Campo Bagatin
\and
	S. Eggl
\and
	W. Losert
\and
	N. Murdoch
\and
	E. Opsomer
\and
	F. Radjai
\and
	D. C. Richardson
\and
	M. Salazar
\and
	D. J. Scheeres
\and
	S. Schwartz
\and
	N. Taberlet
\and
	H. Yano
}


\institute{D. Hestroffer \at
              IMCCE, Paris Observatory, universit\'e PSL, CNRS, Sorbonne Universit\'e, Univ. Lille, F-75014 Paris, France\\
              Tel.: +33 1 4051 2260\\
              \email{daniel.hestroffer@obspm.fr}           %  \\
\and
	P. S{\'a}nchez \at
	CCAR, University of Colorado Boulder, Boulder, CO 80309, USA
\and
	L. Staron \at
	Inst. Jean Le Rond d'Alembert, Sorbonne Universit\'e, CNRS, F-75005 Paris, France 
\and
	A. Campo Bagatin \at
	Departamento de F{\'i}sica, Ingenier{\'i}a de Sistemas y Teor{\'i}a de la Se{\~n}al, Universidad de Alicante, E-03080 Alicante, Spain\\
	Instituto Universitario de F{\'i}sica Aplicada a las Ciencias y las Tecnolog{\'i}as, Universidad de Alicante, E-03080 Alicante, Spain
\and
	S. Eggl \at
        IMCCE, Paris Observatory, universit\'e PSL, CNRS, Sorbonne Universit\'e, Univ. Lille, F-75014 Paris, France\\
        LSST / DiRAC Institute, Department of Astronomy, University of Washington, Seattle, WA 98105, USA
\and
	W. Losert \at
	Institute for Physical Science and Technology, and Department of Physics, University of Maryland, College Park, MD 20742, USA
\and
	N. Murdoch \at
    DEOS/SSPA, Institut Sup\'{e}rieur de l'A\'{e}ronautique et de l'Espace (ISAE-SUPAERO), Universit\'{e} de Toulouse, 31400 Toulouse, France
\and
	E. Opsomer \at
	GRASP, Research Unit CESAM, University of Li\`ege, B-4000 Li\`ege, Belgium
\and
	F. Radjai \at
	LMGC, Universit\'e de Montpellier, CNRS, F-34090 Montpellier, France
\and
	D. C. Richardson \at
	Department of Astronomy, University of Maryland, College Park, MD 20742, USA
\and
	M. Salazar \at
	LICB, Univ. de Bourgogne, CNRS, F-21078 Dijon, France
\and
	D. J. Scheeres \at
	Aerospace Engineering Sciences, University of Colorado Boulder, Boulder, CO 80309, USA
\and
	S. Schwartz \at
	Labratoire Lagrange, Univ. Nice, CNRS, Observatoire de la C{\^o}te d'Azur, F-06304 Nice, France\\
    Lunar and Planetary Laboratory, University of Arizona, Tucson, AZ, USA
\and
	N. Taberlet \at
	Universit{\'e} de Lyon, Laboratoire de Physique, {\'E}cole Normale Sup{\'e}rieure de Lyon, CNRS, F-69364 Lyon, France
\and
	H. Yano \at
	JAXA/Institute of Space and Astronautical Science, 3-1-1 Yoshinodai, Chuo-ku, Sagamihara, Kanagawa  252-5210, Japan
}

\date{Received: date / Accepted: date}

\maketitle

\begin{abstract}
Asteroids and other Small Solar System Bodies (SSSBs) are of high general and scientific interest in many aspects. The origin, formation, and evolution of our Solar System (and other planetary systems) can be better understood by analysing the constitution and physical properties of small bodies in the Solar System. Currently, two space missions (Hayabusa 2, OSIRIS-REx) have recently arrived at their respective targets and will bring a sample of the asteroids back to Earth. Other small body missions have also been selected by, or proposed to, space agencies. The threat posed to our planet by Near-Earth Objects (NEOs) is also considered at the international level, and this has prompted dedicated research on possible mitigation techniques. The DART mission, for example, will test the kinetic impact technique. Even ideas for industrial exploitation have risen during the last years. Lastly, the origin of water and life on Earth appears to be connected to asteroids. Hence, future space mission projects will undoubtedly target some asteroids or other SSSBs. 

In all these cases and research topics, specific knowledge of the structure and mechanical behaviour of the surface as well as the bulk of those celestial bodies is crucial. In contrast to large telluric planets and dwarf planets, a large proportion of such small bodies  is believed to consist of gravitational aggregates (`rubble piles') with no---or low---internal cohesion, with varying macro-porosity and surface properties (from smooth regolith covered terrain, to very rough collection of boulders), and varying topography (craters, depressions, ridges). Bodies with such structure can sustain some plastic deformation without being disrupted in contrast to the classical visco-elastic models that are generally valid for planets, dwarf planets, and large satellites. These SSSBs are hence better described through granular mechanics theories, which have been a subject of intense theoretical, experimental, and numerical research over the last four decades.

This being the case, it has been necessary to use the theoretical, numerical and experimental tools developed within Soil Mechanics, Granular Dynamics, Celestial Mechanics, Chemistry, Condensed Matter Physics, Planetary and Computer Sciences, to name the main ones, in order to understand the data collected and analysed by observational astronomy (visible, thermal, and radio), and different space missions.

In this paper, we present a review of the multi-disciplinary research carried out by these different scientific communities in an effort to study SSSBs.

%Include keywords, PACS and mathematical subject classification numbers as needed.
% \PACS{PACS 02.70 \and 06.60\and 45.20\and 45.70\and 46.55\and 81.05\and 83.10\and 95.10\and 95.75\and 96.25\and 96.30}
\end{abstract}

%%
%%_________________________________________________________________________________________________________________________

\section{Introduction}
\label{S:1}
 
Small Solar System Bodies (SSSBs) correspond to a class of Solar System objects, all orbiting around the Sun, that are smaller than dwarf planets and are likely not internally differentiated. With the exception of active comets---that has bright and large comas---the SSSBs appeared stellar-like (not resolved) in the 19th century telescopes, thus explaining the origin of the `asteroid' designation. This naming and classification has evolved with time, between minor planet and asteroid designation, to end up in 2006, with the definition of the SSSB class, following an IAU resolution \citep{iauXXVI2008}. So, Small Solar System Bodies currently include most of the asteroids, Trans-Neptunian Objects (TNOs) and comets. SSSBs are hence found in different regions of the Solar System: Near Earth Objects (NEOs) orbiting in the vicinity of the Earth (with perihelion $q \le 1.3\,$AU), Main Belt Asteroids (MBAs) orbiting between Mars and Jupiter, Trojan asteroids co-orbiting Jupiter or other major planets, Centaurs orbiting between Jupiter and Neptune, and in the belt of trans-Neptunian objects (TNOs) beyond the orbit of Neptune (with semi-major axis $a \ge 30\,$AU). Additionally, and because they share some properties with these small bodies, we shall also consider in the rest of the paper some small planetary satellites (orbiting around a planet, and hence not strictly speaking SSSBs). 

These small icy or rocky bodies are of particular interest for fundamental scientific research to understand the formation and evolution of planetary systems in general, and of our Solar System in particular. SSSBs, being pristine in general, or having experienced little geological evolution, are valuable tracers of the early time of the Solar System.  
Asteroids, comets and planetesimals are also believed to be the source of water on Earth through past collisions \citep[and references therein]{morbidelli2000,obrien2014,bancelin2017} and so they have been proposed to be the fundamental bricks that formed an environment apt to support life on Earth. 
Space missions are of great value to deepen our understanding of SSSBs, as will be seen in Sect.~\ref{S:2}. Following the success of the Hayabusa space mission that visited the asteroid (25143) Itokawa, other sample-return missions (Hayabusa 2, OSIRIS-REx) are currently underway and will greatly enhance our scientific knowledge about asteroids. 
The Martians Moons eXplorer (MMX) will target small planetary satellites as part of a sample return mission, the missions Lucy and Psyche have recently been approved to target a metallic asteroid and several Jupiter Trojans, respectively, and other missions have also been proposed to space agencies (MarcoPolo-R, Phobos-Grunt-2, ZhengHe, Castalia, DePhine, OKEANOS, ...).

On a more societal aspect, there are a number of asteroids (potentially hazardous asteroids - PHAs) with a non-zero probability of impacting the Earth.  These impacts can yield extinction level events at planetary scales on astronomical/geological time frames of several tens of million of years, or more local natural disasters on a hundred-years time-frame \citep[and references therein]{board2010,pelton2015}. 
Understanding our vulnerability to such PHAs (for life and goods) necessitates a detailed understanding of the results of an impact on the ground (land or water), the physics of the entry in the atmosphere, and the structure and properties of the asteroid itself. 
In addition to the DART space mission, several missions have been proposed to assess the mitigation techniques and capabilities, including their dependence to the asteroid's properties (Don Quijote, AIDA with DART and its complementary part HERA, NEOTwIST).

The interest in asteroids has also increased during the last decade for economic reasons, in exploration and exploitation. Near Earth Asteroids were the intended target for the next human exploration, or resource collection by a redirect mission (the Asteroid Redirect Mission, ARM, currently dismissed) before considering a human mission to Mars. Extraction of possible extra-terrestrial resources in space (In-Space or In-Situ Resource Utilisation - ISRU) is raising interest from industries for business purposes, mostly for use in space and for sustaining habited missions in space. This is pushing governments to adopt laws in agreement with the Outer Space Treaty\footnote{\url{http://www.unoosa.org/oosa/en/ourwork/spacelaw/treaties/introouterspacetreaty.html}} of the UNO-SO.  This also requires a thorough understanding of the targeted object and moreover, where and how to mine potential resources \citep{galache2017}. As a consequence of these interests, several space missions have targeted, or are proposed to target, asteroids, comets, and satellites; with a prevalence to Near Earth asteroids - because of their proximity to Earth (and therefore reduced mission duration and costs).  Main belt asteroids are mostly seen as an opportunity fly-by during a cruise to a planet.

Presently, more than 700,000 asteroids have been identified and discovery is still progressing with current surveys (for instance, the discovery rate is of more than 1800/year for only NEOs \citep{chamberlin2018}), and more is to come as the limiting magnitude of these surveys is pushed to track fainter objects. The recent WISE/NEOWISE space mission \citep{mainzer2014} has revised downward the estimation of the total population of asteroids by size range; however, a large fraction of  objects smaller than approximately 100 meters still remains to be discovered. The whole mass of TNOs is estimated to be at least one order of magnitude larger than that of asteroids, but they are very faint and few details on their structure are known -- except in the case of the dwarf planet Pluto. Therefore, most of the current discussion in this paper will be focused on asteroids. 

There is no strict classification of objects by size, in the size-range of several meters to hundreds of kilometres.   
Even if small in size, compared to planets or giant satellites, asteroids can no longer be considered as tiny point-like masses and have gained interest as small worlds on their own right. 
Moreover, asteroids are driven by a great variety of both dynamical and physical mechanisms and show a large variety and diversity of composition, size, shapes, morphology, and surface properties according to what is observed. Indeed, asteroid orbital dynamics are mainly driven by their interaction with the Sun through gravity and other gravitaional or non-gravitaional perturbations.  In this respect, they are considered as test particles for some General Relativity tests \citep{will2014}.  However, given the accuracy reached in many studies on their dynamics (either short- or long-term) it appears that the effects connected to their physical properties can no longer be neglected. The size and shape have a direct effect on astrometric measurements, or photocentre offset to the centre of mass, and subsequently on their orbit determination or improvement \citep{hestroffer1998}. Mutual perturbations between asteroids or by telluric planets are affecting their orbit propagation \citep[e.g.][]{hilton2002, mouret2007}. Even without considering such secondary effects, the mass and density are  fundamental parameters that characterise an asteroid. After the discovery of binary asteroids in the early 1990s, such as (243) Ida-Dactyl \citep{chapman1995} from space, and (45) Eugenia from ground-based telescope \citep{merline1999}, and some space probe visits to asteroids, as the recent mission to (25143) Itokawa \citep{fujiwara2006}, mass and porosities\footnote{More precisely the macroscopic porosity.} could be estimated, completely renewing our interpretation of these objects from a geological point of view. 

Many asteroids and small bodies are supposed to be `rubble-piles' (a terminology introduced by C.R. \citet{chapman1977} for asteroids shattered as the result of impacts, and gravitationally reaccumulated into a single body) and not monolithic bodies governed only by material strength. This term of rubble-pile is however confusing due to conflict with the different structural interpretation commonly used in Geology. For this reason, other terms have been proposed: gravitational aggregates, self-gravitating aggregates and, more recently, granular asteroids. These low bulk density asteroids are thus gravitational aggregates held together mostly by mutual gravity \citep{richardson2002_ast3, bagatin2001}, with surfaces covered by ponds, craters, grooves, boulders, large topographic slopes, etc. For instance, the 17\,km size asteroid (433) Eros shows a large crater and a smooth and deep surface of regolith, while (25143) Itokawa, approximately 500\,m large, presents a shallow layer of gravel-like grains. All these observations have revealed a number of surface features that are yet to be fully understood \citep{murdoch2015_ast4}. 
In contrast to large gas giants or planetary bodies that are generally in hydrostatic equilibrium, these smaller bodies are able to sustain comparatively large shear stresses, yields, and plastic deformation. Their surfaces and interiors as well can be modelled as granular media formed by particles ranging from dust to boulders. So, during the last two decades, asteroids have often been modelled as granular systems, with many works from different groups and authors; a summary can be found in a few review papers \citep[][and references therein]{richardson2002_ast3, murdoch2015_ast4, scheeres2015_ast4, hestroffer2017, bagatin2018}. However, a comprehensive theory of granular media is elusive, depending on a large set of parameters and not easily scalable to planetary objects.

In the following sections we provide a review of the concept of granular systems, its application to small bodies in planetary science, and connected open questions. We start with Sect.~\ref{S:2} that presents the current knowledge on the bodies under study, what has been learned from observations and the models that have been developed. Section~\ref{S:3} gives a general introduction on the theory of grains and granular systems, and what should be relevant for planetary science of gravitational aggregate objects. Section~\ref{S:4} will present the general modelling techniques to study granular media as either a continuum (FEM) or as a discrete (DEM) system. In Sect.~\ref{S:5} we explore experiments on granular systems either on the ground (that is under Earth's gravity), or in a micro-gravity environment. After presenting benchmark cases in Sect.~\ref{S:6}, and applications to the study of cohesion, spin-up and YORP effect, segregation, and post-impact re-accumulation, we give a general perspective in Sect.~\ref{S:7}.

%%
%%_________________________________________________________________________________________________________________________
\input{AARv_NUMAGS_S2}
%\section{Asteroids and small bodies}

%%
%%_________________________________________________________________________________________________________________________
\input{AARv_NUMAGS_S3}
%\section{Granular systems and Granular mechanics}

%%
%%_________________________________________________________________________________________________________________________
\input{AARv_NUMAGS_S4}

%\section{Physical Models}

%%
%%_________________________________________________________________________________________________________________________
\input{AARv_NUMAGS_S5}
%\section{Experiments}

%%
%%_________________________________________________________________________________________________________________________
\input{AARv_NUMAGS_S6}

%\section{Benchmarking cases}

%%
%%_________________________________________________________________________________________________________________________
\input{AARv_NUMAGS_S7}

%\section{Discussions and prospective}

%%
%%_________________________________________________________________________________________________________________________
\paragraph{\it Additional notes:}

Since the elaboration and submission of this paper manuscript, two exploration and sample-return missions,Hayabusa2 (from JAXA) and OSIRIS-REx (from NASA), have arrived to their respective targets. While more results will be published in the future, both target asteroids appear to be gravitational aggregates.

In June of 2018 the Hayabusa2 spacecraft arrived at asteroid (162173) Ryugu and begin its exploration and characterisation of that body \cite{watanabe2019}. In August the spacecraft made a descent close to the surface to measure the total gravitational attraction of that body which, when combined with its shape determination yielded a bulk density of 1190\,kg/m$^3$. The surface of Ryugu was rocky enough so that the planned touchdown sampling was delayed until February 2019, however the spacecraft successfully made contact with the surface and exercised its sampling procedure at that time.  

In December 2018 the OSIRIS-REx spacecraft arrived at asteroid (101955) Bennu and began to characterise that body \cite{lauretta2019}. The overall bulk density of the asteroid was found to be 1190\,kg/m$^3$, the same value as Ryugu, and both of their macro-porosities were estimated to be up to 50\%, consistent with being a `rubble pile' \cite{scheeres2019bennu,watanabe2019}. The surface of both Bennu and Ryugu were seen to be uniformly covered by boulders across a large size distribution, with very few regions covered by finer regolith \cite{walsh2019,sugita2019}. At a morphological level, both Ryugu and Bennu have many similarities, and thus future comparisons between the bodies will be of great interest.

%%
%%_________________________________________________________________________________________________________________________
\begin{acknowledgements}
% * <hestro@imcce.fr> 2018-05-23T14:20:59.968Z:
% 
% ALL: please complete list of acknowledgements
% 
% ^ <hestro@imcce.fr> 2018-05-23T14:21:45.185Z.
This work is a direct result of support by the International Space Science Institute,
ISSI Bern, Switzerland, through the hosting and provision of financial support for the international team ``Asteroids \& Self Gravitating Bodies as Granular Systems'' led by DH. The authors would like to thank the ISSI Institute and staff for their support, and the Paris observatory for financial support. Thanks to MIAPP, Munich Institute for Astro and Particle Physics of the DFG cluster of excellence ``Origin and Structure of the Universe'' and participants of the the workshop on NEOS for fruitful discussions.
EO thanks Prodex (Belspo) and ESA (Topical Team No. 4000103461) for financial support. DCR was supported in part by NASA grant NNX15AH90G awarded by the Solar System Workings program.
SRS acknowledges support from the Academies of Excellence: Complex systems and Space, environment, risk, and resilience, part of the IDEX JEDI of the Universit\'e C\^ote d'Azur. SE acknowledges support from the DiRAC Institute in the Department of Astronomy at the University of Washington. The DIRAC Institute is supported through generous gifts from the Charles and Lisa Simonyi Fund for Arts and Sciences, and the Washington Research Foundation.
\\
We are grateful to all the other members of the ISSI international team for discussions, exchanges, inputs, and contributions. We are grateful to Brian Warner for kindly providing us an up-to-date `spin-rate versus diameter' figure.
\\
This work has made use of Wm R. Johnston archive data \url{http://www.johnstonsarchive.net}, and intensive use of NASA's Astrophysics Data System.
\end{acknowledgements}

%\begin{verbatim}
%as required. Don't forget to give each section
%and subsection a unique label (see Sect.~\ref{s:1}).
%
%Text with citations \cite{RefB} and \cite{RefJ}.
%
%\paragraph{Paragraph headings} Use paragraph headings as needed.
%\begin{equation}
%\end{equation}
%
%\end{verbatim}
%--------------------------------------------------------

\bibliographystyle{spbasic}      % basic style, author-year citations
\bibliography{AARv_NUMAGS_ref}   % name your BibTeX data base

\end{document}

%% file: AARv_NUMAGS_S2.tex
\section{Asteroids and small bodies}
\label{S:2}

\subsection{Observations and knowledge}
\label{S:observations}
Asteroids were long suspected to be monolithic, possibly differentiated bodies, dry and dense, in particular in the inner regions of the Solar System, with bare rock surfaces. However, during the last thirty years, our vision of asteroids' surfaces and interiors has considerably evolved. This is a result of the particular advancement in observational techniques (from the ground and from space) that has allowed us to remotely characterise a number of SSSBs.  Over time, this has resulted in a significant progress of our knowledge and understanding of asteroids, satellites and comets, which has increased even further with the analysis of the data collected through in-situ space exploration.  Since the first review book on asteroids \citep{chapman1977} we have not only increased, but deepened our knowledge of their physical characteristics (sizes, shapes and masses, composition, thermal inertia, albedos, etc., and their statistics).  Astronomers and planetary scientists have also discovered new features, objects or phenomena (Yarkovsky, YORP, planetary migration, binaries and multiple systems, tumblers, fast rotators, main-belt comets MBC and active asteroids, to mention a few) which were described in the subsequent books Asteroids II, III, and IV \citep{1979ast,1989ast2,2002ast3,2015ast4}. The same holds true for comets \citep{wilkening1982comets, festou2004comets2}, though to a lesser degree.  
 
\subsubsection{Asteroids and collisions} 
The population of asteroids originates from steady-state collisional dynamics \citep{dohnanyi1969}, so that the number density distribution, for a mass $m$, follows a power law, with:
\begin{equation}
  N (m) \propto m^{-11/6}
  \label{E:sizedist}
\end{equation} 
\noindent  with the smaller objects being more numerous. C. R. Chapman proposed that---as a result of the inter-collisional evolution---larger asteroids would be fractured or constitute a recollection of fragments that did not escape after the catastrophic collision; he introduced the terminology of `rubble pile' \citep{chapman1977} (see Sect.~\ref{S:1}). Such bodies would hence be self-gravitating collections of blocks---or gravitational aggregates---with more or less fractured rocks, with no or little internal cohesion, sometimes highly fractured and porous bodies, with more or less coherent arrangements of the constituting blocks. This vision has been confirmed with the estimation of porosity \citep{britt2001}, see Eq.(~\ref{E:porosity}), for some of the asteroids for which the bulk density could be measured. Space missions to small bodies  (see Table~\ref{T:missions}), either as flybys or rendez-vous, have considerably improved our knowledge of these bodies, by bringing detailed measurements and observations of their interiors and surfaces. Flybys are valuable even if they are only brief encounters, unfortunately, however, they are not systematically incorporated in mission trajectories. 

The NASA NEAR mission showed that the bare rock surface predictions for (433) Eros were incorrect.
 Indeed, this in-situ mission---{followed by other} space missions---revealed a substantial layer of unconsolidated rocky material and dust (regolith) covering the surface of (951) Gaspra, (243) Ida, (433) Eros, (21) Lutetia, and (25143) Itokawa \citep{murdoch2015_ast4}.
Additionally, this regolith seems to be active: there is evidence of motion in the form of landslides, particle migration, particle size sorting and regolith production. So, the geophysical characteristics of the surfaces of small planetary bodies appear to be diverse and complex \citep{murdoch2015_ast4}. Also, observations made by the Hubble Space Telescope of the disruption events of active asteroids P2013/R3 \citep{jewitt2014} and P2013/P5 \citep{jewitt2013} imply that not only asteroid surfaces are covered by regolith, but that their internal structure is granular and not monolithic. 

The usual assumption---that prevailed for more than a century---that asteroids are single monoliths, is also incorrect. While long speculated in the end of last century \citep{weiden1989_ast2}, and albeit negative result from surveys around the largest minor planets \citep{gehrels1987}, evidence has been obtained only in the last decades that ``binary asteroids {\it do} exist'' \citep{merline2002_ast3}. It is noteworthy that there are no satellites orbiting around the largest asteroids of the main belt (the dwarf planets), but satellites and moons are found in different asteroid dynamical classes (Near Earth Objects, main belt asteroids, Trojans, Centaurs, trans-Neptunian Objects, ...). Furthermore, some have rings \citep{braga2014, ortiz2017}. The number of detected gravitationally bounded binaries and multiple systems, and also asteroid pairs, has rapidly increased, showing also a higher proportion of systems in the NEO and TNO populations; though -- due to observational bias -- they are not all sampled identically. The formation scenario of these binary or multiple systems, containing a variety of mass fraction, separation distances, angular momentum values, etc. is not fully understood. Different mechanisms can be required within various populations of asteroids. For examples, formation mechanisms can include fission and mass shedding, post catastrophic collision, re-accumulation, capture, ... \citep{noll2008_ssbn, margot2015_ast4, walsh2015_ast4, tardivel2018}. 

\subsubsection{Spin rate, size, and shape} 
\begin{figure}
  \includegraphics[width=\textwidth]{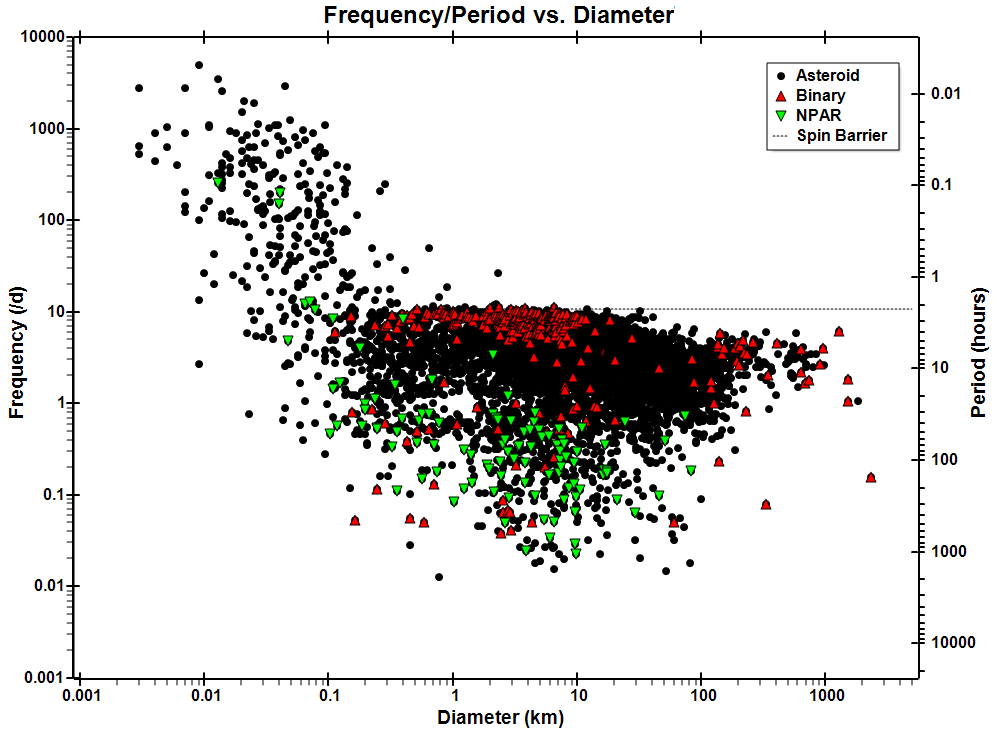}
\caption{Spin period for more than 5500 asteroids as function of their estimated diameter. A `spin barrier,' above which no large objects is rotating, is apparent. This is consistent with a  gravitational aggregate structure of loose conglomerations held together by mutual gravitation, yet it is due mainly to a gravity dominated regime. Conversely, small objects---in the strength dominated regime---can spin very fast, with full rotation within minutes; a possible hint towards a more monolithic structure, or presence of additional cohesive strength. Near Earth objects are predominant in the population of objects smaller than approximately 1\,km. (Source: the {\tt LCBD} light curve database \citep{warner2009}).}
\label{fig:lcdb} 
\end{figure}

There are several fundamental parameters to characterise asteroids, both physical and dynamical. One such fundamental parameter is the spin-rate or period of rotation, that can be obtained relatively easily together with the object's brightness. The spin period is obtained with some confidence from the light-curve (for those objects that are not spheroidal in shape); while the size is generally estimated from the brightness, sometimes by assuming the object's albedo (when no direct or radiometric measure is available). This provides information for a spin-rate versus size plot, which gives a good view of some of the general properties of the objects we are studying (Fig. \ref{fig:lcdb}). In particular, two features are apparent: 1) a spin barrier at about $P \approx 2.4\,$hr for the larger objects, and 2) the existence of fast and superfast rotators \citep{pravec2000,pravec2002_ast3,scheeres2015_ast4}. 
The spin barrier was supposed to be good evidence that most large objects in the range $1-100\,$km -- in particular close to spin limit -- are granular in nature or `rubble-piles', and not monolithic \citep{bagatin2001}, but this is merely the fact that such large bodies are in a gravity dominated regime \citep{scheeres2015_ast4} which does not necessarily imply cohesionless aggregates. Fast rotators on the other hand, must be either monolithic rocks with natural strength, or have additional cohesion to bound a granular structure (see Sect.~\ref{S:4}); such differences are however not recognisable from simply the size and spin-rate parameters.

When these bodies reach their critical spin rate, which is dictated by their size, density and internal strength and strength distribution, they will undergo deformation, surface mass shedding, fission or catastrophic disruption events \citep{hirabayashi2015, sanchez2016, sanchez2018}. Particles that are ejected near the equatorial plane can supposedly re-accumulate and form moonlet(s) \citep{walsh2008_nat}. A more abrupt fission event could also produce a binary system \citep{tardivel2018} or could be at the origin of the observed asteroid pairs \citep{pravec2010,sanchez2016,zhang2018}. Only the structurally strongest asteroids will reach the 2.4\,h spin barrier.
More generally, whatever the spin-rate is, the surface gravity on a body of several $100\,$m size is in any case weak (approx. $10^{-3} - 10^{-6}\, g$), and escape velocities are in the cm/s range.

Our modelling of asteroid shapes has dramatically evolved during the last decades. The observed periodic variation in an asteroid lightcurve over a rotation period can be explained by either a variation of the projected surface for a non-spherical or non-spheroidal shape, or by a variation in albedo of the observed surface \citep{russell1906}. It eventually appears that such variations are mainly driven by shape effects \citep{kaasalainen2001a, kaasalainen2001b, li2015_ast4}. Starting from simple tri-axial ellipsoid models spinning about their shortest axis (as an outcome of collisional and dynamical processes), or more sophisticated cellinoids \citep{lu2016}, we now have tools to derive convex and non-convex topographic shapes \citep{kaasalainen2002_ast3}. Stellar occultations and high angular resolution observations valuably complete that information and modelling \citep{durech2015_ast4}, as well as radiometric and polarimetric observations for size and albedo \citep[and references therein]{masiero2017, stansberry2008_ssbn, harris2002_ast3, delbo2015_ast4, belskaya2015_ast4}.
Radar observations---by analysing the echoed wavefront on the asteroid's surface---also provide information on shape and more detailed physical properties \citep{ostro2002_ast3, benner2015_ast4}.  
The body's shape and outer envelope can be related to its internal structure, and also modify its spin barrier level \citep{harris1996,holsapple2001}.

\subsubsection{Mass, bulk-density, and porosity}
Binaries and multiple systems are of particular interest since, as a result of collisions, they could correspond to the case proposed by \citet{chapman1977} and likely be gravitational aggregates. Moreover, binaries are of particular interest here, because by deriving their mutual orbits, it is possible to derive another fundamental parameter: the (total) mass of the asteroidal system. 

\begin{figure}
  \includegraphics[width=\textwidth]{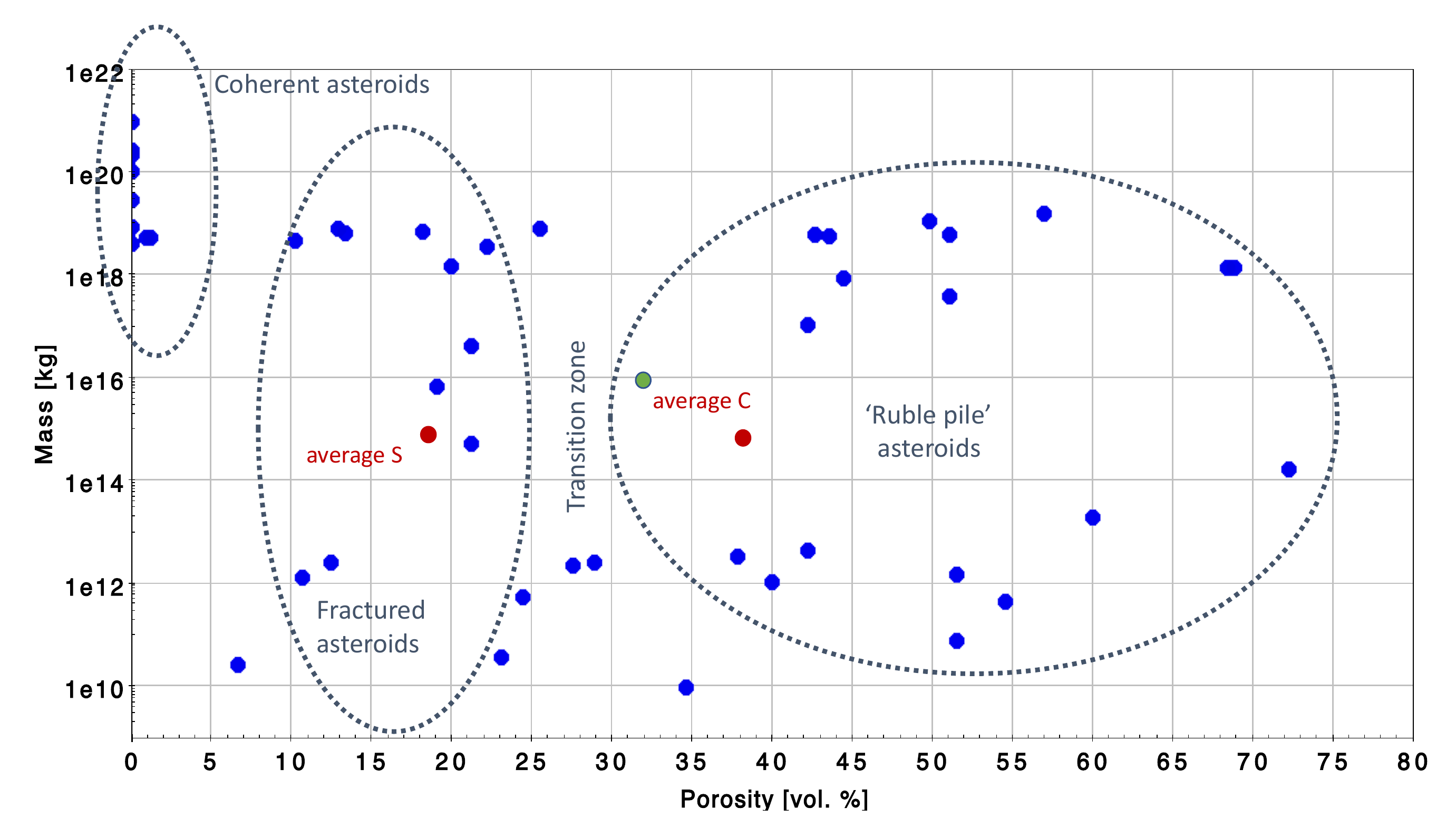}
\caption{Possible internal structure arrangement of gravitational aggregates, based on their macro-porosity (adapted from \citet{britt2002_ast3,carry2012}). Note that the uncertainty on the porosity is generally large ($\pm 5$\% and more).}
\label{fig:porosity}  
\end{figure}

Measuring the mass is a difficult goal to reach which can be achieved from the careful astrometric observations of binary systems, or otherwise from the analysis of their gravitational perturbation during a rendez-vous or a fly-by with a space probe, or a close encounter with another (small, target) asteroid. Such close encounters between asteroids are much more frequent than space mission fly-bys, but far less precise or accurate, and require high accuracy astrometric measurement together with a global inversion to take into account all the effects \citep{mouret2007,baer2017}. 
In any case, estimating the mass of any small planetary object studied is of high importance  
as it allows an indication of its density, porosity, internal structure or internal mass distribution, and its global behaviour to stresses or impacts. 
Given the bulk density $\rho$ of a small body (generally measured from the knowledge of mass and volume, with some uncertainty), and $\rho_g$ the mean density of the material that constitutes the constituting grains (generally unknown, and estimated from the taxonomic class), the macro-porosity\footnote{Micro-porosity on the other hand is in the matrix of the grains or meteorites. Micro-porosity is a porosity that will survive entry in the atmosphere.} $p$ is defined by:
\begin{equation}
  p = 1-\rho_g/\rho
  \label{E:porosity}
\end{equation}
\noindent It is a scale invariant parameter that also measures the intersticial voids in the global structure of the body. The bulk density uncertainty is dominated by the error on the volume  which can be large. Besides, the porosity estimation depends on our knowledge of the interior material or internal constituents, and the densities of analogue meteorites (i.e., those representing a good match to the asteroid taxonomic class). Progress in the classification, and determination of such meteorite densities and micro-porosities has been achieved in laboratories. However, the meteorites collected on Earth may not always be the best analogue for representing an asteroid as a whole, as it could also be formed of other materials, containing volatiles or fragile material, or having regions with different porosities. This means that our knowledge on asteroids' porosity remains limited to a few objects, and with substantial uncertainty (see Fig. \ref{fig:porosity}). 

\begin{figure}
\centering  \includegraphics[width=0.8\textwidth]{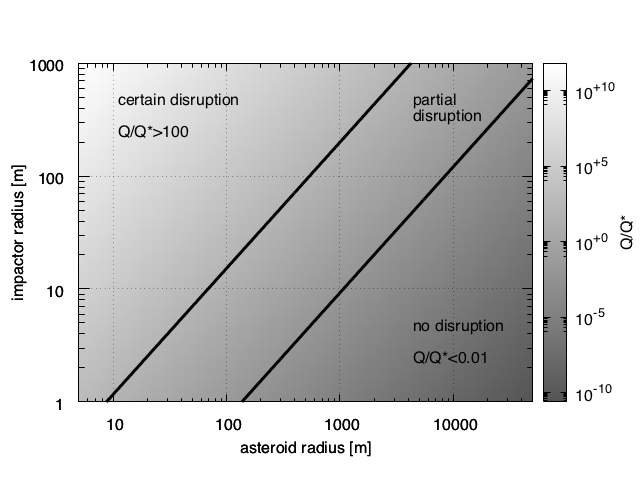}
\caption{Disruption limits for collisions as a function of impactor  and target size. 
The relative velocity at impact is 5\,km/s, typical for main belt collisions. The target asteroids are assumed to be non-porous and spherical ($\mu=0.55$) with a density of 3500\,kg/m$^{3}$.}
\label{fig:asteroid_fracture}      
\end{figure}

Porosity and material strength play crucial roles in how asteroids react to impacts of smaller debris or "projectiles" as they determine how shock-waves and damage zones caused by the disruption event propagate through the target \citep[e.g.][]{asphaug_1998, jutzi_2010,brucksyal_2013}. Several studies have found, for instance, that strength dominated asteroids below a few hundred meters in size are more difficult to disrupt if they are porous \citep[][and references therein]{jutzi_2014}. For larger asteroids self-gravity starts to dominate, and the role of porosity becomes less clear cut. Cases where a collision leads to an asteroid's disruption has been discussed extensively in literature \citep[see e.g.][]{henych_2018}. Comparing the imparted kinetic energy per unit mass $Q$  to  the specific energy needed to disrupt a target $Q^*$ can provide useful first insights. Let \citep{henych_2018}: 
\begin{eqnarray}
Q &=& \frac{1}{2} \frac{m\; u^2}{M}, \cr\cr
Q^*&=& (44 R^{-0.6\mu} +68 R^{3\mu})\;[\cos(\phi)\;U]^{2-3\mu},
\end{eqnarray}
where $m$ is the mass of the projectile, $M$ the mass of the asteroid, $R [km]$ the radius of the asteroid in kilometers, $u [m/s]$ the relative impact velocity in $m/s$, and $U$ the same quantity in km/s. Furthermore, $\phi$ denotes the impact angle and $\mu$ is the so-called ``point-source scaling-law exponent'' that is around $\mu=0.4$ for porous materials and $\mu=0.55$ for non-porous rock-like material.  
Both, $Q$ and $Q^*$ are given in units of J/kg. The ratio between $Q/Q^*$ determines whether or not enough energy has been deposited to disrupt the target. A value of $Q/Q^* \approx 1$ characterises an impact that dismantles about 50\% of the body.  A value much greater than unity indicates pulverisation. On the other hand, $Q/Q^* \ll 1$ describes a simple cratering event. Fig. \ref{fig:asteroid_fracture} shows that typical main belt collision speeds suffice for relatively small "projectiles" to affect the structural integrity of kilometer-sized asteroids. Catastrophic disruption is more likely to occur in collisions between main belt asteroids of similar size.

The result of a post-catastrophic re-accumulation should be a loosely bound gravitational aggregate \citep[e.g.][]{tanga2009, michel2013}. Surprisingly, low bulk densities of the order of $1200-1500\,$kg/m$^3$ were observed on some asteroids, either single or binaries (e.g. Mathilde, Eugenia, ...), while chondritic or silicate material densities are of the order of $2500-3500\,$kg/m$^3$.  This is supposed to be evidence of the presence of large voids in the interior of the body, that can be connected to macro porosity of the order of $40-50\,$\%. How force chains and friction are acting, or how the blocks are arranged inside of the body, is not clear. The mechanism that prevents the fine regolith grains at the smooth upper surface from filling the voids in the case of porous bodies could be linked to friction \citep{britt2001}, but needs to be better understood with regard to seismic shaking and segregation mechanism. How asteroids react to collisions isn't clear neither; collisions could either increase the porosity, voids, and fragments, or otherwise reduce the porosity through compaction. The asteroid (253) Mathilde is an example of an object with very high macro-porosity with smooth surface and large craters indicating that it has survived major impacts.  However, the actual internal structure, or the size distribution of its constituent particles remains unknown.

\begin{figure}
  \includegraphics[width=10cm]{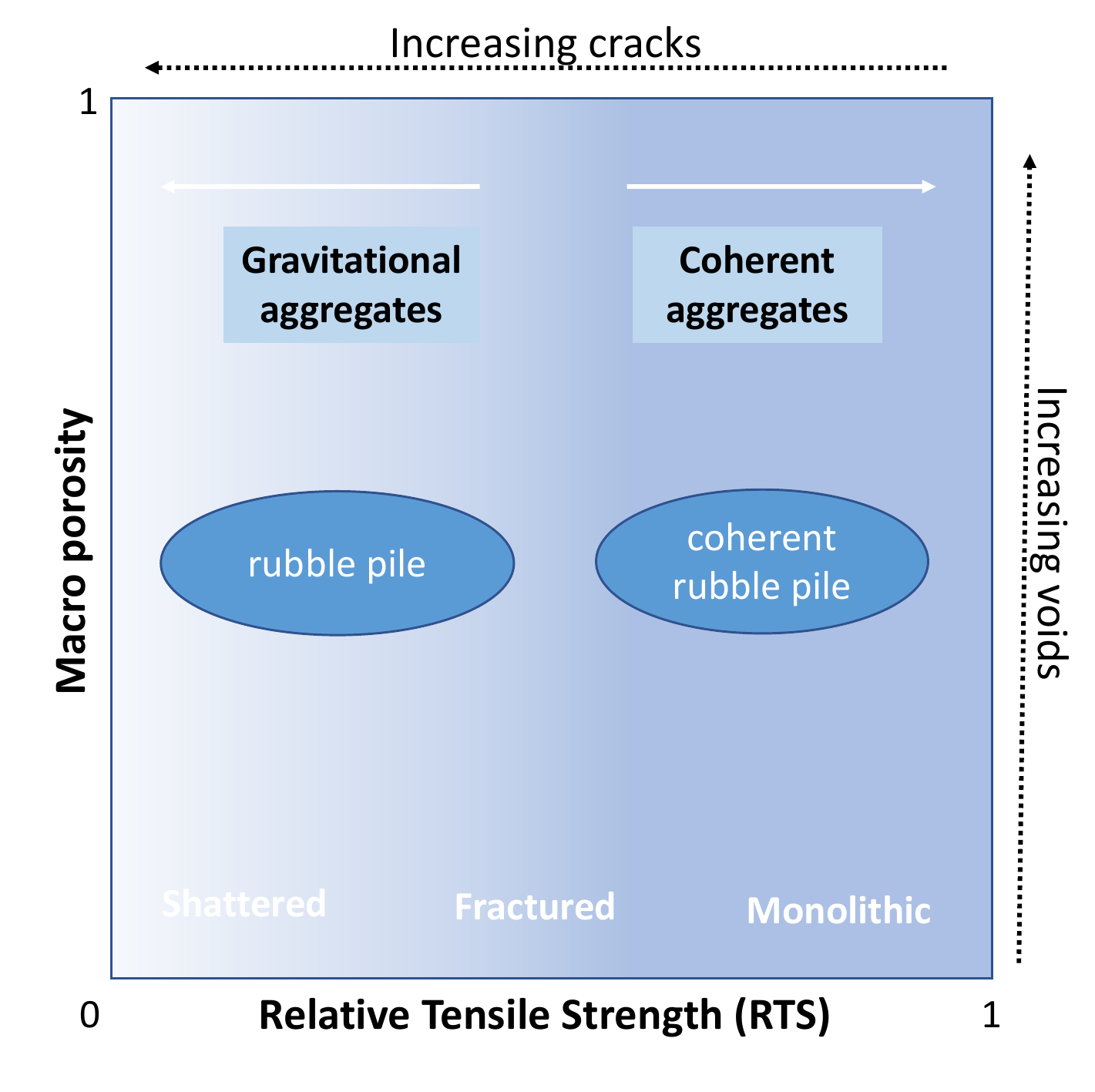}
\caption{Relative tensile strength (RTS) and porosity parameter space (adapted from \citet{richardson2002_ast3}).}
\label{fig:GA-Richardson}
\end{figure}

From the knowledge of asteroids mass and porosity, \citet{britt2002_ast3} identified three classes of objects that could reflect different internal structures: coherent, fractured, and loosely consolidated (see Fig.~\ref{fig:porosity}). Further, \citet{richardson2002_ast3} schematically characterised gravitational aggregates by relative tensile strength (RTS), in addition to macro-porosity (see Fig.~\ref{fig:GA-Richardson}). These are driving parameters that qualitatively identify different classes of internal structures and global mechanical behaviour. 
So, there could be different regimes between strength dominated and gravity dominated bodies \citep[see][]{holsapple2002_ast3}, with different reactions to mechanical tensile or compressive stresses and impacts. Thus, bodies in the range of concern in this paper are in the transition regime between purely strength dominated (like stones and meteorites) and gravity dominated coherent bodies (like major planets), and can react differently to impacts and cratering \citep{asphaug2002_ast3,britt2002_ast3}. For instance, porous and structurally weak bodies may, through stress dissipation, be highly resistant to catastrophic disruption. Additionally, it modifies the cumulative size-distribution slope from Eq.~(\ref{E:sizedist}) \cite[][and references therein]{bottke2005}.
Lastly, all the mechanisms present in granular media can govern gravitational aggregates' shapes, strength and failure limits \citep{scheeres2015_ast4} (impacts, landslides, fission, tides, ...).

\begin{table}
\caption{List of flown space missions to small bodies (\citep[updated]{hestroffer2017}).  Missions to the dwarf planets Ceres, Vesta, and Pluto, or to large planetary satellites, have not been included here; while two launched and ongoing---not yet achieved---sample-return missions have been included. These are mainly NASA, JAXA, and ESA space agencies interplanetary missions, with contributions from national agencies, CNSA, and former USSR for the `Halley armada' and Phobos2. Legend for mission type: F=fly-by ; O=orbit or/and hovering ; SR=sample return ; L=landing ; I=impact ; Date is given at arrival.}
\label{T:missions}
\begin{tabular}{llll}
\hline\noalign{\smallskip}

\rowcolor{DarkRed}
\color{white}{\bf Target} & \color{white}{\bf Mission} & \color{white}{\bf Type} & \color{white}{\bf Date} \vspace{1mm}\\

\rowcolor{Gray}
{\bf Asteroids}&&&\\
(101955) Bennu & {\it OSIRIS-REx} & O+SR & {\it Dec. 2018} \\
(162173) Ryugu & {\it Hayabusa2} & O+SR & {\it June 2018} \\
(4179) Toutatis & Chang'E 2 & F & Dec. 2012\\
\rowcolor{Gray}
(21) Lutetia & Rosetta & F & July 2010\\
(2867) Steins & Rosetta & F & Sep. 2008\\
\rowcolor{Gray}
(25143) Itokawa & Hayabusa & O+SR & June 2005\\
(5535) AnneFrank & Stardust & F & Nov. 2002\\
\rowcolor{Gray}
(433) Eros & NEAR Shoemaker & O+L & Feb. 2000\\
(2685) Masursky & Cassini/Huygens & F & Jan. 2000\\
\rowcolor{Gray}
(9969) Braille & Deep Space 1 & F & Jan. 1999\\
(253) Mathilde & NEAR Shoemaker & F & June 1997\\
\rowcolor{Gray}
(243) Ida+Dactyl & Galileo & F & Aug. 1993\\
(951) Gaspra & Galileo & F & Oct. 1991\\
\hline
\rowcolor{Gray}
{\bf Comets}&&&\\
\begin{tabular}{l}
67P/Churyumov-\\
Gerasimenko\\
\end{tabular}
& Rosetta & O+L & Aug. 2014 \\
\rowcolor{Gray}
103P/Hartley 2 & EPOXI (Deep Impact ext.) & F & Nov. 2010\\
9P/Tempel 1 & Deep Impact, Stardust & O+I & July 2005, 2011\\
\rowcolor{Gray}
81P/Wild 2 & Stardust & O+SR & Jan. 2004\\
19P/Borrelly & Deep Space 1 & F & Sep. 2001\\
\rowcolor{Gray}
Grigg-Skjellerup & Giotto & F & July 1992 \\
1P/Halley & 
\begin{tabular}{l}
The Halley armada: \\
Vega 1\&2, Sakigake,\\
Suisei, Giotto, ICE\\
\end{tabular}
& F & March 1986\\
\rowcolor{Gray}
21P/Giacobini-Zinner & ICE & F & Sep. 1985\\
\hline
{\bf Moons}&&&\\
\rowcolor{Gray}
M1 Phobos &
\begin{tabular}{l}
MEX, ODY, MGS, \\
Viking1 + Phobos2 \\
\end{tabular}
& F+O &
\begin{tabular}{l}
2004, 2001, 1997,\\
1989, 1976
\end{tabular}\\
M2 Deimos & 
MEX, ODY, MGS, Viking2 
& F & 2004, 2001, 1997, 1976\\
\rowcolor{Gray}
satellites of Saturn & Cassini & O & 2000-2017\\
\noalign{\smallskip}\hline
\end{tabular}
\end{table}

\subsection{Links between asteroids, planetary satellites, and comets}
\label{S:links}

Comets and asteroids show very distinctive physical and dynamical properties. On the dynamical side, first, objects with parabolic or hyperbolic orbits, i.e. eccentricities $e \ge 1$, would be categorized as comets (or interstellar objects \citep{meech2017}). Next, the Tisserand parameter $T_J$ for a body whose periodic orbit is characterised by a semi-major axis $a$, eccentricity $e$, and inclination $i$:
\begin{equation}
  T_J = {a_J \over a} + 2\,\sqrt{{a \over a_J}\,(1-e^2)}\,\cos i
\end{equation}
\noindent given here with respect to the massive Jupiter (with orbit $a_J \approx 5.2\,$AU), is a well known dynamical parameter that broadly separates objects that are strongly perturbed by Jupiter, or not, as a function of their encounter velocity. This helps to separate the orbital classes between asteroids (typically $T_J>3$), and comets (typically $T_J<2$ for Halley type comets - HTC, and $2<T_J<3$ for Jupiter family comets - JFC). On the physical and visible side, when approaching the Sun at approximately less than 3\,AU, a comet begins to sublimate its volatiles, giving rise to a large and bright coma and tail, making comets very distinct in appearance to asteroids. 

Now, extinct comets, after having lost their volatiles and after a dynamical evolution bringing them in the Near-Earth orbiting region, are difficult to distinguish from asteroids using remote observations: \cite[``a comet in disguise"][]{kerr1985,weissman2002_ast3}. On the other hand, the old viewpoint that asteroids---as a kind of minor planet---are dry with no volatiles, possibly differentiated with iron core and mantle and hence dense, while all objects beyond the snow line at approximately 5\,AU are icy bodies that have retained their volatiles, has changed. Indeed, `active asteroids' (and not only the large dwarf planet Ceres) or `main belt comets' have been observed in the main belt, well below the hypothesised snow line \cite[``a comet among the asteroids"][]{hsieh2006, jewitt2015_ast4}. Moreover, as difficult as it is, ices have been detected on the surface of a few asteroids \citep{rivkin2010,campins2010}, and it has been shown that volatiles can have long lifetimes if buried under a moderate layer of regolith \cite[][and reference therein]{schorghofer2008,delbo2015_ast4}. So, volatiles can be present in asteroids as it is the case in comets, albeit in different proportions. Conversely, some Centaurs have shown activity at large distance from the Sun, invalidating the commonly accepted scenario of possible activity \citep{jewitt2009}. Moreover, in-situ collection of comet grains by the Stardust mission has revealed the presence of silicates, so that comets are not only made of volatiles and interstellar dust but share also some composition with---closer to Sun---asteroids. This has been confirmed by the Rosetta mission, which gives more insight into the grains and dust, composed of compact grains and fluffy aggregates \citep{aclr2015}, rich in carbon and non-hydrated minerals \citep{bardyn2017}. This is supported by our current understanding of the formation and dynamical evolution of the whole Solar System, showing a big mixing of the distant and inner regions \citep{demeo2015_ast4} with conglomerates that include refractive material and volatiles, hence erasing or fading any initial solar nebula density or composition gradient. Lastly, the derived masses of some asteroids show very low bulk densities, which is assumed to reflect a high porosity (Mathilde density of $\approx 1.3$ \citep{yeomans1997}, Eugenia density of $\approx 1.2$ \citep{merline1999}), similarly to comets. The same low densities have been found for Phobos and Deimos (with values of 1.5 and 1.9, respectively \citep{rosenblatt2016}), for Trojans and TNOs (Patroclus density of $\approx 0.8$ \citep{marchis2006}), and other extremely low densities are found in comets and small Saturnian moons \citep{thomas2010}. 

This all has pushed a change of paradigm in the last decades in our understanding of the real differences between the different classes of minor bodies, bringing forth the concept of an asteroid-comet continuum \citep{bockelee2015}. Additionally, the irregular satellites of Jupiter, Saturn, Uranus, Neptune, the small moons of Pluto, the moons of Mars Phobos and Deimos, etc., have been hypothesized to be either captured asteroids or TNOs \citep[][and references therein]{peale2015}. Such capture origin remains sometimes unclear, and  impact scenarios are also possible as in the case of Mars. Nevertheless---whatever their origin---the similarities in size, surface and composition remain. 
Thus asteroids, comets and small moons show common features. Although they have some variation and gradients, for instance in the constituent ingredients or even with different formation mechanisms, their final global structure and general mechanical behaviour could still be modelled with a similar approach.

\subsection{Summary and Open questions}
\label{S:openQ}

We have a general knowledge of the surface composition of these small bodies, from telescopic observations, and sometimes of their surface roughness through thermal inertia \citep{delbo2015_ast4,mueller2017}. But detailed knowledge on the presence of regolith, craters, boulders and `chaotic terrain', or other geological features is much more limited. And in that case, they have shown a large diversity that is still difficult to predict, or to unequivocally correlate with observable data such as size, taxonomic type and spin-rate. 
We also expect to see segregation phenomena acting at the surface of these bodies. However size or density segregation and sorting in granular media, and moreover, how it really behaves on self-gravitating small granular bodies, is still a matter of research.
Exploration of these small bodies has now entered a new era with landers and sample return devices that are designed to make contact with their surfaces.  This was started with the Hayabusa and Rosetta missions, and we are awaiting the event---and success---of both Hayabusa-2 and OSIRIS-REx missions. In all these cases, or in the case of the DART kinetic impactor on Didymos' secondary, different mechanisms and scenarios have been considered and the actual reaction and behaviour of the surface will be of high interest for our general understanding. Many space missions are motivated for scientific, or exploration reasons, often requiring a contact with the surface, or landing and manoeuvring. Response of the surface during anchoring, drilling, or sampling is of the utmost importance and needs a better understanding of the physics of granular media in low gravitational conditions \citep{daniels2013}. The same applies for mitigation in space of a Potentially Hazardous Asteroid (PHA) that would likely impact the Earth. In that case, the threatening asteroid's material properties can play a fundamental role when changing its trajectory via impulsive deflection techniques, or when attempting its complete disruption \citep{ahrens1992,sanchez_2009,sugimoto_2014, eggl_2015}.

As seen before, detailed knowledge on asteroids' surface composition and terrain has been obtained for only a small sample of targets,  but still nothing is certain about their interior.
Tomography \citep{herique2017}, seismic investigation \citep{murdoch2017_pss} (either passive or active), in addition to classical planetary geodesy \citep{yeomans2000, konopliv2002, abe2006, rosenblatt2008, andert2010,jacobson2007, paetzold2017, mcmahon2018}, and possibly with CubeSats \citep{walker2016,murdoch2016_egu,hestro2017}, adapted to low-gravity small bodies and their perturbed environments have been proposed as viable techniques to probe the interior, and internal structure of relatively small asteroids.
At present, most of the available evidence on the internal composition of small solar system bodies is indirect: bulk-density, rotation periods, and crater sizes. As said before, it has been observed that several asteroid classes can be defined in the asteroidal population depending on their bulk porosity, reflecting different degrees of porosity \citep{britt2002_ast3}. At the moment, the origin of such porosity is not completely understood, yet we think that collisions could likely have played a role. However, this alone does not explain for instance the differences between asteroids Mathilde and Itokawa which have similar bulk densities, but different surface characteristics.

Given these observational facts, it is evident that we need to better understand the possible evolution of asteroids: how gravitational aggregates evolve under the YORP effect, how this depends on size/mass, on internal cohesion or other contact forces. What is the formation mechanism of binaries and multiple systems: continuous mass shedding, fission, or reaccumulation? How strong are these objects to meteoritic bombardment and tides, kinetic impactor, sampling mechanism, etc.? Given the large variety and diversity among these objects we should limit our assumptions, as we often need to span large parameter sets. Most of these require theoretical and numerical modelling, benchmarking, and experiments.  This is the focus of the following sections.

%% file: AARv_NUMAGS_S3.tex
\section{Granular systems and Granular mechanics}
\label{S:3}

\subsection{What are grains? }
\label{S:whatGrains}

A grain is a discrete, rigid, macroscopic particle that can interact with other particles through dissipative contacts, and the motion of which can be accurately described by Newtonian dynamics\footnote{Possibly including relativistic effects, but this is not relevant in this paper on small bodies.}. This means that the behaviour of the grains is dominated by particle interactions and gravity forces, in comparison to which thermodynamic agitation appears negligible \citep{jaeger_1996}. 
The behaviour of an assembly of grains without cohesion or adhesive forces is  dictated by the interactions at the individual grain scale. In dense granular systems, aside
from the gravitational attraction, the interactions are essentially mechanical contact forces, which can be decomposed into normal forces (arising from the elasto-plastic behavior of the material) and tangential forces (due to the frictional properties of the grains).

Although grains found in nature and asteroids can display a variety of shapes, the case study of spherical grains has been at the focus of theoretical and experimental studies in view of shedding light on the mechanical properties of generic granular materials. 
A good approximation of the  repulsive force involved when two rigid spherical grains---assumed to be identical---are in contact,  is given by the Hertz Law~\citep{landau1986}: 
\begin{equation}
F_{\mbox{Hertz}} = E \frac{\sqrt{2R}}{3 ( 1 - \nu^2)} \delta^{3/2}
\label{E:hertz}
\end{equation}
where $R$ is the radius of the grains, $E$ is Young's modulus and $\nu$ is Poisson's ratio of the material, and $\delta = 2R - r$ (where $r$ is the distance between the centres of the grains) is the overlap that represents the elastic deflection at the contact area \citep{Agnolin2007c}. 
The details of the calculation are somewhat accessory but the striking result is the non-linearity of the interaction, since the repulsive force scales as the $3/2$ power of the overlap. This specific power arises from the spherical shape of grains, although the calculation is based on the theory of linear elasticity. However, the nonlinear nature of the grain-grain interactions  remains valid for a wide variety of particle shapes and for large contact deformations.   

In addition to the elastic component of the normal force, there exist dissipative effects during particle collisions. Energy dissipation can originate from the visco-elastic properties of grains as well as their plasticity, causing irreversible deformations at  the contact scale,  which  result in inelastic collisions and solid friction \citep{andreotti2013}. Friction is a fundamental aspect as it leads to an indeterminacy problem.  Indeed,  Coulomb's law for solid friction sets only bounds on the value of the tangential forces between two solid bodies. Hence, the internal mechanical state of an assembly of grains is not uniquely determined by the position and velocity of the grains, but may display residual stresses which induce strongly history-dependent mechanical responses \citep{toiya2004,Agnolin2007a}. 

Due to the rigidity of the individual grains, granular packings may sustain intense forces, which propagate into the system through force chains. Force chains are a manifestation of the high heterogeneity of granular systems in general, and explain why the mean stress state in a granular system does not fully describe the actual mechanical behaviour exhibited by the system \citep{Liu1995a,Jaeger1996c,Radjai1996,Radjai2015}. 

On Earth, other effects include cohesion forces caused by capillary bridges due to the humidity, air drag acting on particles or thermal agitation \citep{RadjaiRichefeu2009,andreotti2013}. Due to the absence of atmosphere, these effects appear to be negligible in small asteroids.   At the same time, the extremely weak gravitational environment will make other forces, particularly van der Waals cohesion and electrostatic forces, appear comparatively stronger and therefore, important in the behaviour of the system. On asteroids, electrostatic forces, which are typically of the order of $10^{-6}\,$ N, can easily exceed the gravitational forces in a micro-gravity environment. However, an appropriate modelling of such forces still remains challenging \citep{hartzell2017}.

\subsection{What are granular systems? }
\label{S:whatGranular}

A granular system is a collection of grains that interact through binary or multiple contact interactions, and whose mean behaviour results from the collective dynamics of the grains. 
  Due to the complexity and richness of the grain-scale interactions, granular systems display a wide variety of behaviours (thixotropic solid, liquid, gas), depending on the external mechanical excitation or loading~\citep{Jaeger1996c,richard2005}. 
 
In the presence of a large input of energy in the system (for instance, large slope angles, vibrating boundary conditions), grains form dilute assemblies which are reminiscent of molecular gases. Binary collisions occur between grains whose mean free path is larger than their typical diameter. Examples of such gaseous systems include some of the less dense planetary rings, in which individual grains interact through rare collisions.  In less energetic settings, granular systems flow in dense packings dominated by long-lasting multiple contact interactions, and frictional dissipation. Rock avalanches or landslides as observed on Earth and Mars \citep{lucchitta1979,neuffer2006} mostly fall into this category. In this case, the mean behaviour is reminiscent of that of a viscous fluid, however with strongly non-Newtonian properties. Finally, at equilibrium or in quasi-static flow,  a collection of  grains may resist stress without immediate deformation, a behaviour characteristic of a solid \citep{liu1998,GaoZhaoLiEtAl2014}. `Rubble-pile' asteroids fall in this category and may undergo gravitational, tidal or centrifugal forces without rearranging. 

It appears therefore vain to propose a general theoretical framework for the description of such a wide range of behaviours. Instead, specific theories have been developed, or adapted, for each specific state. Kinetic theory (based on the classical thermodynamics molecular kinetic theory)  has proven to be very successful in modelling and describing the physics of dilute assemblies of grains \citep{jenkins1985,brilliantov2010}. On the other hand, classical soil-mechanics and structural engineering theories  are often applied to describe the mechanical properties of dense granular material under slow deformation  \citep{terzaghi1996,mccarthy1977}. Finally, hydraulic equations (or the more complex Navier-Stokes equation) are applied to describe the liquid state of granular matter,  coupled with semi-empirical models for the effective viscosity \citep{midi2004}. These models will be presented in greater details in Sect.~\ref{S:4}.

\subsection{Phenomenology}
\label{S:phenomenology}
When a flow starts, it will do so for slope angles lower than the angle at which avalanching started, thereby  implying a modification of its frictional properties, often referred to as the hysteretic behaviour of the granular pile. 
While flowing, the internal friction will be affected by the flow dynamics and the pressure, and thus the context of the flow itself will strongly affect the properties of the granular mass \citep{midi2004}. The characterisation of the internal friction of model granular media has thus prompted many research activities. A significant amount of progress has been made, but there are still many aspects relevant to asteroids to explore, including cohesive forces and grain size segregation.
A model for the granular rheology in the simple case of mono-sized non-cohesive grains  will be presented in Sect.~\ref{S:4}.
  
 Another important aspect of granular behaviour in the context of small solar system bodies is their ability to segregate grains according to their size. This phenomenon is omnipresent on Earth, in geophysical flows (debris and rocks flows) which classically involve grain sizes covering three to four orders of magnitude, but also in human activities (food industry, civil engineering, powder technologies etc). When a granular bed is submitted to vibrations or is flowing, larger grains and smaller grains tend to separate, with the larger grains classically rising to the surface  of the granular system (also the potential level on Earth). Essentially, this phenomenon results from the higher probability that smaller grains have to fill in the gaps opening in the system while flowing or shaking, a mechanism known as the kinetic sieve \citep{savage1988}. Size segregation results in patterns that may partly affect the properties of the system. In the case of unconfined flows for instance (like rock flows), larger boulders are segregated at the front and pushed aside by the flow, forming channels or lev\'ees  which will in turn affect the flow dynamics \citep{felix2004}. In the context of small solar system bodies,  understanding the dynamics of segregation may allow for the interpretation of grain size distribution at the surface of asteroids in terms of formation history, structure, exposure to impacts or other sources of agitation. Although much studied, the general formulation of a lift force that would describe the segregation dynamics is still mostly lacking \citep{guillard2014}. Presenting existing models is beyond the scope of the present paper, however, interesting insights will be found in \citet{kudrolli2004}.

\subsection{What is relevant?}
\label{S:whatRelevant}
%[Paul]

The previous sections have clearly outlined  an introduction to the study of granular matter, granular media, the intricacies of some puzzling phenomena observed in nature and shed some light on our understanding of their inner workings.  The following sections on the other hand, will provide the reader with a more in-depth, theoretically strict and sound explanation of the current theoretical, experimental and numerical methods that have been developed and used to study granular systems.  However, as insightful as this all is, we need to answer two very fundamental questions, as otherwise, all this knowledge will simply be inapplicable: 
\\
\indent 1. are small Solar System bodies granular in nature? and if so, 
\\
\indent 2. how does all we know about granular dynamics relate to them? 
\\

The answer to the first question came as a result of the space missions that visited these bodies (see Table \ref{T:missions}).  As seen in Sect.~\ref{S:2}, snapshot images taken of those asteroids showed them to be covered with surface regolith, to be heavily cratered, and in the case of Ida to have a binary asteroid companion which itself had a regolith covering. Since these initial observations, the space-based and ground-based observations of asteroids have grown along with the realisation that these bodies are fundamentally granular in nature and hence that their physical evolution must be described using principles of granular mechanics applied to these extreme environments, which brings us to answer the second question.

Since other planetary bodies apart from the Earth have been out of human reach for most of our history, most of our knowledge of granular systems is restricted to systems on Earth.  What this means is that they were all subjected to an almost constant gravitational field with a magnitude of $\approx 9.81\,$m.s$^{-2}$, in which cohesive and electrostatic forces were negligible, except for fine powders (with diameters $\approx 10^{-6}\,$m). The accumulated knowledge collected by early builders \citep{fall2014}, engineers, scientists \citep{faraday1831} and the common farmer had these two premises that were true for any practical purpose, but not so much for asteroids, comets, or even the Moon.  Additionally, on Earth grains always interact with fluids \citep{burtally2002,kok2012}, be this in the form of wind (and atmosphere) or water currents (and humidity) in most cases.

Given the size and shape differences between the Earth and small planetary bodies, the gravity vector on their surfaces is not as strong as on Earth and changes dramatically from point to point.  This is compounded by the rotational state of these bodies, and the possible influence of solar wind and solar radiation pressure.

In spite of these environmental differences, phenomena such as size segregation, phase transitions and granular flows have either been hinted \citep{sierks2011} or observed directly \citep{jewitt2013} and this is precisely the point: these are only environmental differences.  Fundamentally, each individual grain is still solid, grain-grain interactions are still governed by surface, contact forces, these interactions are still very dissipative, and individual grains still follow the laws of Newtonian physics.  Therefore, there is no fundamental difference between granular systems on Earth and those in SSSBs; not even for two solitary grains colliding in the vast emptiness of space.  In that sense, a diffuse planetary ring  \cite[e.g.][]{wisdom1988,salo2001,tiscareno2018} could be seen as a self-gravitating granular gas, whereas SSSBs could be seen as self-gravitating granular systems in a condensed, solid phase which will be the focus of the rest of this paper.  That being the case, the theoretical, numerical and experimental tools and methods that have been already developed should still be adequate to study them as long as we allow for the implementation of a correct gravitational field, and cohesive and electrostatic forces at the bare minimum.  As \citet{scheeres2010} point out, cohesive and electrostatic forces become important due to the greatly reduced magnitude of the gravitational field in comparison to Earth's gravity.

This is the approach that many scientists in the Planetary Sciences and Granular Dynamics community have taken.  Numerical tools need to implement particle-particle cohesive and electrostatic forces as well as self-gravity in the case of discrete element methods (DEM) \citep{richardson2000,stadel2001,sanchez2011,richardson2011}.  Whereas in the case of finite element methods (FEM) \citep{hirabayashi2016, hirabayashi2014,holsapple2004}, the specific geometry of the gravitational field is calculated from the specific shape of the studied body, with a given rheology and equation of state of the material.  Experiments need to be carried out in droptowers \citep{sunday2016}, and in aeroplanes that perform a number of parabolic flights to emulate lower gravitational conditions (milli-g levels for tens of seconds are common) \citep{murdoch2013}.  In the best case scenario, experiments can be carried out in the International Space Station as this guarantees micro-gravity conditions for extended periods of time \citep{fries2016,fries2018}. The following sections in this paper will detail how this has been carried out, the caveats and main results.

%% file: AARv_NUMAGS_S4.tex
\section{Physical Models}
\label{S:4}

\subsection{Continuum models}
\label{S:continnum}

Granular systems are by definition discrete, namely made of a multitude of individual macroscopic components whose state can be individually described in classical terms: position, velocities, contact and volume forces. However, when zooming out and observing the motion of a granular system as a whole (flowing for instance), a mean behaviour emerges from the multiple interactions of the individual grains: they can be described as an equivalent continuum media, with effective mean properties. Nevertheless, because grain interactions are dissipative in nature, and energy is constantly lost within the granular mass, the theoretical step necessary to derive continuum equations and properties from individual grain-scale behaviour is far from straightforward.  As was described in Sect.~\ref{S:3}, depending on the external energy available to the system, a granular mass can exist in a dilute state and behave like a gas, or flow in a dense state and resemble a viscous fluid, and eventually stop and become like a solid with specific elasto-plastic properties. Each of these three states of granular matter can borrow the theoretical tools classically used in thermodynamics, fluids mechanics  and soil mechanics respectively. We will provide a short introduction to each of them in the following. However, in nature, granular systems very quickly transit from one state to the other: a granular flow down a steep slope will transform into a gas  but will soon stop solid on a gentle topography, while a static granular heap may start deforming as the result of a quake or any other perturbation, leading to shear banding (namely failure) or rapid flow.  The physical understanding and modelling of these transitions remain a challenge, and little will be said here on these aspects. 

\subsubsection{Granular gases}
\label{S:gas}

When a granular mass is very dilute, grains interact through short-lived binary collisions, in between which each experiences  an independent free flight over a distance comparable to their diameter. In this situation, they resemble the molecules of a gas. Because grains are macroscopic, they do not exhibit a temperature in the thermodynamical sense. However, by analogy with molecular gas, where temperature describes the fluctuating part of the kinetic energy, one may introduce a granular temperature as the mean grain's velocity fluctuations \citep{ogawa1978,haff1983,jenkins1983}: 
\begin{equation}
T = \langle \delta v ^2 \rangle =   \langle (v - V)^2 \rangle, 
\label{eq:T}
\end{equation}
where $v$ and $V$ are the grain's individual velocity and the mean velocity respectively. The definition of a granular temperature lays the basis for the derivation of a kinetic theory of granular gases, by analogy with molecular gases.  Based on the Boltzmann equation, it implies a complex mathematical formulation that is beyond the scope of  the present paper. Here, we will only introduce how transport coefficients can be derived from simple arguments about the transfer of momentum during collisions. 
Let's consider a collection of grains of diameter $D$  interacting through binary collisions with a typical free-flight distance $\ell$. Between two collisions, grains cover the distance $\ell$ with  a typical velocity given by the velocity fluctuations $\delta v= T^{1/2}$ from Eq.~(\ref{eq:T}), so that the typical collision rate is $T^{1/2}/\ell$. During a collision, grains exchange a momentum of the order of $m \delta v = \rho D^3 \delta v$, where $m$ and $\rho$ are the mass and density of a grain. The pressure in the media, given by the total amount of momentum transfer per unit of time (namely, $( \rho D^3 \delta v)\times (T^{1/2}/\ell)$) over an effective section area $D^2$, will thus obey:
\begin{equation}
P \simeq \rho\, {D \over \ell}\, T
\label{eq:P}
\end{equation}
and we recover the proportionality between pressure and temperature observed in molecular gases.

When the granular gas is sheared with a shear stress $\tau$, it deforms with a shear rate $ \dot{\gamma} = du/dz$ accordingly (where $u$ is the velocity in the shear direction), thus defining the viscosity $\eta = \tau/\dot{\gamma}$. The transfer of momentum in the shear direction is thus $m du = \rho D^3 du$ times the collision rate $T^{1/2}/\ell$ over the effective section area $D^2$. Considering moreover that $du/dz \simeq du/D$, since $D$ represents the minimum length scale over which gradients may develop in the media, we obtain for the viscosity:
\begin{equation}
\eta \simeq \rho\, {D^2 \over \ell}\, \sqrt{T}
\label{eq:eta}
\end{equation}

Again, we recover the same proportionality between temperature and viscosity as in molecular gases. Using similar arguments, we can recover the thermal conductivity, which relates temperature gradients to  the heat flux ({\em i.e.} kinetic energy flux), as in molecular gases.

An important aspect of granular gases however is that they lose energy during collisions which are inelastic. Introducing the coefficient of restitution during a collision $e$, the kinetic energy lost per collision is $\Delta E = \rho D^3 (1-e^2) \delta v^2 = \rho D^3 (1-e^2) T$. The rate of dissipation (per unit time and volume) is thus:
\begin{equation}
\Gamma = \rho\, \frac{(1-e^2)}{\ell}\, T^{3/2}
\end{equation}

Associated to conservation equations, these coefficients provide a simple constitutive model to describe granular gases. Moreover, they illustrate the problem posed by granular systems and their rigidity. Indeed, in a granular gas, the free flight distance $\ell $ is not much larger than the grain size and can rapidly vanish if collisions are highly dissipative. In that case, $\ell\rightarrow 0$ and all coefficients diverge. 
More elaborate theories will be found in \citet{andreotti2013}, as well as how kinetic theory has been successfully applied to Saturn's rings, probably the most famous example of granular gas in space \citep{goldreich1978,spahn2006,richardson1994}.

\subsubsection{Soil Mechanics}
\label{S:soil}

At rest, granular systems can sustain shear stresses and remain static, at equilibrium. This is for instance the case when one consider a granular heap, or the face of a dune: part of the weight of the granular material is sustained by the slope, without the latter starting to flow. This is true however only if the slope does not exceed a certain critical value $\theta_c$, which defines the internal angle of friction of the material. Past this value, the material flows until a new equilibrium is reached.          
In a similar fashion, a granular material in a shear cell, where both applied pressure $P$ and shear stress $\tau$ are controlled, will start to deform and flow  only when the shear stress reaches a certain critical value $\tau_c$. This critical value is proportional to the pressure $P$, and the coefficient of proportionality $\mu$ (coefficient of friction) defines the internal friction of the material:
\begin{equation}
\tau_c = \mu\, P = \tan \phi\; P
\end{equation}
\noindent where $\phi$ is the angle of repose defined by Coulomb (1776). This provides at first order a plastic criterion for the failure of granular material: either the shear stress is below the friction threshold and the system remains static, or the shear stress reaches the friction threshold and the system undergoes plastic, irreversible, transformations. %
\begin{figure}
  \includegraphics[width=8cm]{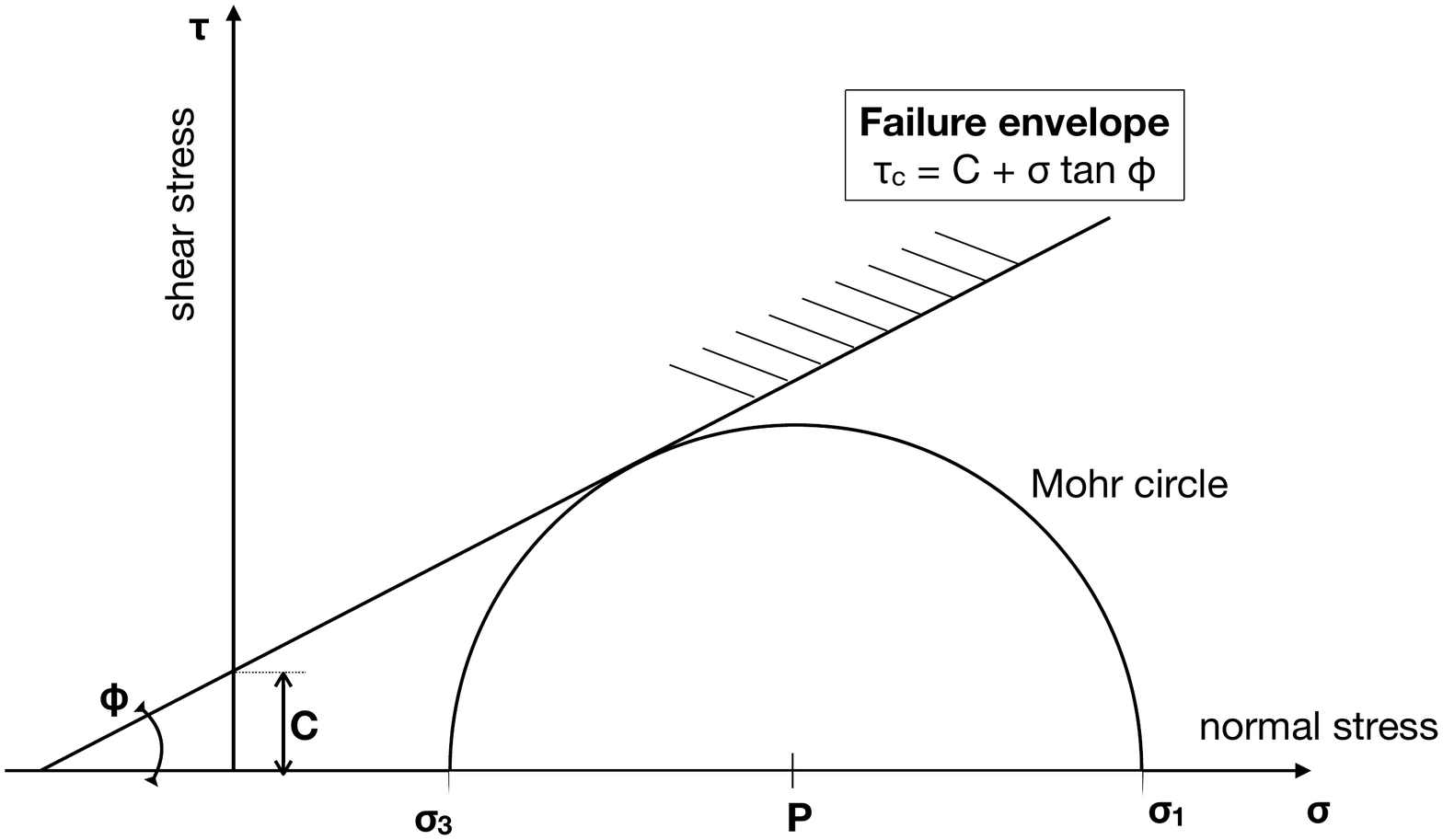}
\caption{The Mohr-Coulomb failure criteria and Mohr's circle. The failure criterion represents the linear envelope that is obtained from a plot of the shear strength $\tau$ of a material versus the applied normal stress $\sigma$, where $\phi$ is the angle of friction, considering also cohesion $c$. }
\label{fig:Mohr}     
\end{figure}

This is  the basis of the Mohr-Coulomb criteria (see Fig.\ref{fig:Mohr}), a well-known tool of soils mechanics, which generalises the above criteria to a fully three-dimensional system for which the direction of the failure is not known a priori. In this case, the ratio $\tau/P$ must be evaluated in all possible directions. The derivation of Mohr's circle implies no specific subtleties: it is recovered by projecting the forces applied to a given surface as a function of the surface inclination, and the details of the calculation will be found in \citet{andreotti2013}. It provides a graphic way of understanding the distance of a given stress state to the material plastic limit. The existence of cohesive forces between the grains is easily taken into account  by the addition of an effective cohesion $C$ to the critical stress $\tau_c$, which is then defined as follows:
\begin{equation}
\tau_c = C + \mu\, P
\end{equation}
\noindent so that, in contrast to a purely loosely bound media, the higher the cohesion, the higher the stress that can be sustained before failure of the granular material. The static angle of repose is the maximum slope that can be supported before the formation of an avalanche, and a dynamic angle of repose is the slope that results after this avalanche has taken place.
However, for granular matter, a resisting difficulty lies in the unambiguous definition of the frictional properties, as explained in Sect.~\ref{S:phenomenology} above. Moreover, granular matter behaves differently if densely or loosely packed, and probably also in different gravity regimes. This difference of behaviour is specifically important when a packing is subjected to a shear stress. If initially densely packed, a granular packing will need to dilate to adapt to the shear. This will lead to the localisation of the deformation and the occurrence of shear banding. By contrast, an initially loose packing will adapt shear stresses by deforming in an homogeneous fashion. Hence, a given granular material will exhibit a different flow rule depending on its initial state. Critical state theories address these aspects, relating the system distance to the plastic limit ({\em i.e.} to failure) to its volume fraction \citep{roux1998}. 

Finite element methods (FEM) have been used in this context to see how friction can account for the global shapes and stabilities of small bodies, or derive stresses and yields in their interior, and possible failures or surface shedding conditions. The gravitational field is computed from a specific (and approximated) shape of the body, with a given rheology and possible equation of state of the material. This has been applied to both asteroids \citep{hirabayashi2016, hirabayashi2014d,sharma2013,harris2009,sharma2009,holsapple2004,holsapple2001}, and small planetary satellites \citep{dobrovolskis1982,kay2018}.

\subsubsection{Fluid Mechanics}
\label{S:fluid}

In many natural situations, granular matter flows in a dense state that makes it resemble a liquid. In this case, grains are interacting through simultaneous multiple (rather than binary) collisions with many neighbours, while undergoing large deformation,  so that theoretical frameworks borrowed from molecular thermodynamics or soil mechanics are not relevant. Here, fluid mechanics provides the basic tools, however with many modifications needed. These tools  are essentially the  mass conservation, and the  Navier-Stokes equation which relates the velocity gradients  to the pressure gradients and most importantly, the viscous or frictional dissipation. 

Considering that granular flows, such as rocks or debris flows, are often thin compared to their longitudinal extension, a handy simplification consists in assuming that vertical velocities are negligible, and that the motion of the flow is accurately described by solving the one- or two-dimensional problem along the flow path \citep{savage1989}. In this case, the problem essentially reduces to solving numerically the mass and energy conservation equations which relate flow height and mean flow longitudinal velocity:
\begin{eqnarray}
\nonumber
& &\frac{\partial h}{\partial t} +  \frac{\partial}{\partial x} (h\bar{u})   = 0,\\ 
& &\frac{\partial h \bar{u}}{\partial t} +  \frac{\partial}{\partial x} (h\bar{u^2})   =   \rho\, g\, h\,  \left( \tan\theta -\mu - \frac{\partial h}{\partial x} \right) \, \cos \theta,\label{eq:SW}\\ \nonumber
\end{eqnarray}
where $h$ is the flow height, $ \bar{u}$ the flow longitudinal velocity, $\frac{\partial }{\partial t} $ and $\frac{\partial }{\partial x}$ are the gradients in time and in space, $\rho$ the flow density, $g$ the gravity, $\theta$ the slope of the topography. The main problem to correctly apply the model to real flows is to have an accurate description of the friction coefficient $\mu$, which balances the two driving forces: gravity and pressure gradients.  In depth averaged equations such as Eqs.~(\ref{eq:SW}), the coefficient of friction is a basal term, namely it encompasses all the dissipation that occurs in the bulk and at the interface with the substrate into a single scalar describing frictional interactions at the bottom of the flow. As a result, much effort has been devoted to understanding what controls the value of $\mu$ in real flows \citep{dade1998,kelfoun2011,lucas2014}, on Earth as well as on other planetary bodies.

When the flow configuration is such that the shallow-layer approximation no longer holds (namely when both longitudinal and normal velocities shape the flow), as for instance in vertical collapses or flow through apertures, the full Navier-Stokes equation must be solved, which is challenging in terms of numerical techniques.  An important difficulty is to handle the fact that the flow viscosity may become infinite is some places, where the granular matter turns to a solid-like behaviour, and starts to creep slowly instead of flowing. Here, the viscosity is often defined using the frictional properties of the flow through an effective coefficient of friction $\mu$. The numerical resolution of such complex flows is computationally expensive  and its applicability to natural flows down realistic topographies is limited.
 However, because these more challenging flow situations provide valuable benchmarking cases to test rheological models for viscosity, they have been the subject of much research is recent years \citep{jop2006, pouliquen2006, lagree2011, staron2012, kou2017}. The rheology of granular flows is more specifically introduced in the next section.

\subsubsection{The rheology of granular flows}

As we have seen in previous sections, dense granular matter is in essence frictional and defining a coefficient of friction $\mu$ is both the simplest and the most efficient way to describe it. Moreover, friction is a finite scalar: it does not diverge nor fall to zero, so that constraining its value experimentally seems feasible. A great variety of configurations have been employed to the task, and  the wealth of measurements gathered has eventually led to the identification of a non-dimensional number which controls the value of the friction coefficient of flowing granular matter: the inertial number $I$ \citep{midi2004}. The inertial number can be understood as the ratio of the two characteristic time-scales dominating in a granular flow: the time scale of macroscopic deformation given by the shear rate $\dot{\gamma}$, and the time scale of grain-scale rearrangements under the local pressure $P$. For a collection of grains of diameter $d$ and density $\rho$, 
\begin{equation}
I = \frac{d \mid \dot{\gamma}\mid }{\sqrt{P/\rho}}.
\end{equation}
The dependence of the coefficient of friction $\mu$ on $I$ is phenomenological and implies the knowledge of three parameters: two extremal values $\mu_1$ and $\mu_2$ for the friction, and a coefficient $I_0$ setting how quickly $\mu$ evolves between the two:
\begin{equation}                
\mu(I) = \mu_1 + \frac{\mu_2-\mu_1}{ {I_0 / I} +1}.
\end{equation}
Although the existence of three independent parameters might be seen as limiting the predictive value of the $\mu(I)$ dependence, the latter has proven very efficient in many flow configurations. By definition, friction relates pressure and shear stress. As such, it does not describe the way the system deforms or flows in response to a solicitation, and additional operations are needed to turn it into a full rheological description.   This is done by deriving an effective viscosity $\eta_{\tt eff}$ using the norm of the shear rate and the pressure:
\begin{equation} 
\eta_{\tt eff} = \frac{\mu(I) \mid \dot{\gamma}\mid }{P}.
\end{equation}
This operation is valid only if deformations and shear stresses are aligned, which is satisfied for rapid flows. However, this might not be strictly the case for slow motion verging to creep, for which the $\mu(I)$ rheology may not be sufficient. In these slow regimes moreover, non-local effects may start to play a role, whereby distant shear deformations need to be taken into account to describe the local state. Non-locality is however much beyond the scope of this introduction to granular behaviour; further reading may include \citet{bouzid2015}, and references therein. 

%\subsection{Discrete Element Methods}
\subsection{Discrete, Numerical Models}
\label{S:discrete}

In contrast to continuum models and finite element methods (FEM) in Sect.~\ref{S:continnum} above, discrete methods are modelling granular media as a collection  particles  that conserve some of the geophysical characteristics of the aggregates to be studied and the real particles. Several computer simulations of granular material, structures, and flows have been developed \citep[][and other references in the sections below]{radjai2011, mehta2012}. A discrete element method (DEM) is any of a family of numerical methods that simulates the dynamics of an ensemble of solid, macroscopic, particles.  In these methods, the particles possess not only translational, but also rotational degrees of freedom and the motion of each particle is calculated over time individually, in accordance with its interaction with the rest of the particles in the system.  The particles that constitute the aggregates can be subject in general to a variety of external forces fields (long range) and contact forces (short range).  In particular, for Planetary Sciences applications, these fields are self-gravity (for an asteroid size body), imposed micro- and milli-gravity fields (for asteroid surfaces).   Below, we introduce three of these methods, the first two have already been used for the simulation of SSSBs with mutual gravitational attraction and the third is starting to be used with the same purpose.

\subsubsection{Soft Spheres DEM - SSDEM}
\label{S:soft}

The soft-sphere discrete element (or molecular dynamics, SSDEM) method for simulating granular material consists in applying Newton's laws of motion to individual deformable spheres. 
Each individual grain can be submitted to long-distance forces (most often gravity) and interacts with its neighbours through direct, long-lasting collisions. The method is time-driven, and consists in computing the positions and velocities (both translational and rotational) of each grain at time $t+dt$ based on the state of the system at time $t$. The search for contacts and the integration of the equations of motion can be optimized using the appropriate numerical methods, e.~g. the Verlet~\citep{verlet1967} or linked list~\citep{mattson1999} methods,  and the leapfrog~\citep{hut1995} or the predictor-corrector~\citep{butcher2016} integration schemes, respectively.

Great care must be given to the choice of the force models~\citep{radjai1997}. The normal force acting between two spheres is given by the Hertzian contact (see Eq.~(\ref{E:hertz}) and Sect.~\ref{S:whatGrains}) but a dissipative component needs to be included to model the inelasticity of granular collisions. Most often, the material is considered either visco-elastic (adding a viscous term to the normal force) or plastic (mimicking irreversible deformations of the grains)~\citep{shafer1996}.   In spite of this, many SSDEM codes have implemented a linear spring-dashpot contact law due to its analytical simplicity and the fact that the coefficient of restitution is independent of the impact velocity \citep{herrmann1998}.
The modelling of the frictional tangential forces also deserves attention. Several methods have been proposed, among which the most widely used are the simple regularized Coulomb's law (appropriate for loose systems with no residual stress) and the history-dependent Cundall model~\citep{cundall1979}. 

The main advantage of the soft-sphere discrete element method lies in its truly physical modelling. The forces are computed from solid mechanics and integrated using fundamental laws of physics, while other methods, including event-driven simulations (HSDEM, Sect.~\ref{S:hard}) and cellular automata, may rely on somewhat arbitrary choices. The SSDEM soft-sphere method also has the advantage of computational efficiency, which allows for simulations of a large number of grains (typically up to several millions). 

\subsubsection{Hard Spheres DEM - HSDEM}
\label{S:hard}

In the hard-sphere discrete element method (HSDEM), impacts are treated as instantaneous and point contact, so restitution and momentum conservation equations are used (rather than integrating over the contact duration).  This is appropriate in the ballistic regime, where the time to cover the mean free path ($t_f$) is much greater than time the particles stay in contact during a collision ($t_c$).  This being the physical reality, the approximation that the numerical method makes is to assume that the duration of a collision is exactly equal to zero ($t_c=0$), making the collisions instantaneous and sound speed infinite.  This implies that by construction, all collisions are exclusively binary, and particles cannot sustain long-lasting contacts.  The instantaneity of the collisions is artificial, but the approximation is valid for kinetically active systems.  In the context of granular matter, this method is well suited to the simulation of granular gases and granular flows in a collisional regime, though more generally (and originally) it was used for the simulation of dilute molecular gases \citep{alder1960}.

Numerically, given a set of particles that form the granular system and their dynamical state (position, velocity and forces on them, including e.g. mutual gravity) the method considers that the particles will move ballistically from collision to collision \citep{alder1959}.  This implies that the order in which collisions happen must be calculated in advance, based on the known state of the system.  Every single collision marks an instant at which a pair of particles stops moving ballistically and exchanges momentum.  The method calculates the amount of time that needs to pass between the present instant and all the possible collisions in the system as the particles move ballistically.  Then a list of collisions in hierarchical (chronological) order is build and all particles move for the needed amount time so that the first collision in the list takes place.  At this moment, the system has changed, the collision list has to be rebuilt and the process is indefinitely repeated. A brute force approach to the building of the list would be to update the entire system and, though simple, it is very inefficient for large systems.  To solve that, \citet{lubachevsky1991} suggested to update only the two particles involved in the last collision.

From a computation standpoint, HSDEM poses a challenge for parallel processing, since for maximum realism the particle collisions should be treated in time order, so there is a computational bottleneck as these processes only ever involve 2 particles at a time.  In contrast, in SSDEM, since particle contacts are finite in duration and treated as extra forces in the equations of motion, no bookkeeping is required to treat the collisions in time order, and everything can be parallelised. Both HSDEM and SSDEM need to identify particle neighbours in order to check for potential (HSDEM) or occurring (SSDEM) collisions. The brute-force neighbour search is an intrinsically order $N^2$ calculation; tree codes or other data sorting methods can reduce this to order $N\log(N)$ or better \citep{richardson2011,rocchetti2017}.  This however, implies that some collisions could be missed and so care must be taken to make sure that this does not happen.  Finally, in HSDEM, if the timestep is much less than the relevant dynamical time(s), the trajectories can be treated as linear and any collisions can be predicted in advance; in SSDEM, timesteps need to be small enough to resolve the restoring forces on contact.  This means HSDEM methods are generally much faster than SSDEM in the ballistic regime (e.g. collision and ejectas); the opposite is true in the dense/quasi-static regime, when parallelism wins out despite the tiny steps needed in SSDEM to resolve contact forces.

\subsubsection{Non-Smooth Contact Dynamics CD}
\label{S:ContactDynamics}

The contact dynamics (CD) method is based on a mathematical formulation of nonsmooth dynamics, and the subsequent algorithmic developments by J. J.~Moreau and M.~Jean \citep{Moreau1994,Jean1999, Radjai2009, radjai2011}. As in HSDEM, the particles are assumed to be perfectly rigid (non-deformable), but their contacts can persist in time as in the SSDEM case. 

In the CD method, the rigid-body equations of motion are integrated for all particles using `contact laws' (instead of contact force laws as in SSDEM) expressing mutual exclusion and dry friction between particles. The equations of motion for each particle are formulated as differential inclusions in which possible velocity jumps replace accelerations. The unilateral contact interactions and Coulomb friction law are treated as complementarity relations, or set-valued contact laws. The time-stepping scheme is implicit but requires explicit determination of the contact network. Due to implicit time integration, inherent in the CD method, this scheme is unconditionally stable. At a given step of evolution, all kinematic constraints implied by lasting contacts and friction are simultaneously taken into account---together with the equations of dynamics---in order to determine all velocities and contact forces in the system, by means of an iterative process pertaining to the non-linear Gauss-Seidel method. The latter consists in solving a single contact problem, with other contact forces set to their values from the previous iteration, and iteratively updating the forces and velocities until a convergence criterion is fulfilled. The iterations in a time step are stopped when the calculated contact forces are stable with respect to the update procedure. The convergence criterion is based on the variation of the mean contact force and/or velocity between two successive iterations. In this process, no distinction is made between smooth evolution of a system of rigid particles during one time step and nonsmooth evolution in time due to collisions or dry friction effects. The uniqueness of the solution at each time step is not guaranteed by the CD method, as the particles are perfectly rigid. However, by initializing each step of calculation with the forces calculated in the preceding step, the set of admissible solutions shrinks to fluctuations which are basically below the numerical resolution. In this way, the solution remains close to the present state of forces. When dealing with complex-shaped particles, the same iterative process can be applied although several contact points may occur between two neighbouring particles. The multiple contacts between two particles are treated as independent unilateral constraints.  

The implicit time-stepping scheme makes the method unconditionally stable. Hence, as the small elastic response times are absent from the model, much larger times steps can be used as compared to the SSDEM. The CD method has been extensively employed for the simulation of granular materials in 2D and 3D with various particle shapes. 

\subsubsection{N-body codes, dealing with mutual attraction}
\label{S:nBody}

%An efficient approach is to use a tree code for both the neighbor searching and the gravity calculation.  For gravity, the tree includes moments of the gravity contributions of the particles within a tree cell.  The more moments stored, the better the accuracy, but the more memory and computation time required.  Pkdgrav stores up to hexadecapole order (16-pole) as a compromise between speed and accuracy.  The tree opening angle controls the quality of the calculation: the smaller the angle, the more tree cells are opened, giving better accuracy at increased cost.  Generally the angle should be 0.7 rad or less in Barnes-Hut/k-D-tree style algorithms.  Note that this accuracy criterion is unimportant for neighbor finding, so long as it is not so big as to introduce self-referencing pathologies.  Other approaches include using a grid and cutting off forces from distant particles, but at cost of missing important distant contributions.

Long-range forces such as inter-particle gravity are expensive to compute, since every particle in the system contributes a force on every other particle. There is a great variety of cosmological N-body codes \citep[e.g.][]{klypin2017}, with various approximations used to reduce the force cost. The particle mesh method \citep{hockney1988} can be used to compute the potential in Fourier space, but it is not accurate for small separations. One popular approach is to use a tree code, which assigns particles to cells and approximates forces from distant cells by expanding the gravitational potential around the cell's centre of mass to some fixed order. The computation cost is reduced to $O(N \log (N))$ -- where $N$ is the number of particles in the system -- instead of $O(N^2)$ in the direct pair-wise summation. Various tree algorithms exist, but one of the most widely used is the oct-tree due to \citet{barneshut1986}. Briefly, the space occupied by particles in the system is subdivided recursively into nested cubical cells until each particle uniquely occupies its own cell. To compute the force on a particle, the largest cells are considered first; if a cell subtends an angle less than some critical value (the opening-angle parameter, $\theta$), its contribution to the force is obtained from its moments (see below). Otherwise, its occupied sub-cells are considered recursively in turn until the $\theta$ criterion is satisfied, or the sub-cell contains only a single particle. The speed and accuracy of the method depends on both $\theta$ and the expansion order. As an example, the acceleration up to quadrupole order is given by \citet{richardson1993}:
\begin{equation}
a = -\frac{M}{r^3}\mathbf{r}+\frac{\mathbf{Q\cdot r}}{r^5}-\frac{5}{2}\frac{(\mathbf{r\cdot Q\cdot r})\mathbf{r}}{r^7} ,
\end{equation}
where the gravitational constant $G=1$, $\mathbf{r}$ is the vector between the particle under consideration and the cell's centre of mass, $M$ is the total mass in the cell, and $\mathbf{Q}$ is the quadrupole moment tensor given by
\begin{equation}
Q_{jk} = \sum_i m_i(3x_{i,j}x_{i,k}-r_i^2\delta_{jk}) ,
\end{equation}
where $\mathbf{r}_i=(x_{i,1},x_{i,2},x_{i,3})$ is the position of particle $i$ in the cell relative to the cell's centre of mass, and $\delta_{jk}$ is the Kronecker delta function. Note the quadrupole moment of a cell can be computed recursively from the moments of its sub-cells using the shift theorem \citep{hernquist1987}. Higher order increases the complexity but improves the accuracy for a given $\theta$. A typical goal is to achieve an average force accuracy of 1\% or better, which generally requires at least quadrupole order and $\theta < 0.7\,$rad ($\approx 40\deg$). Other optimizations are possible to improve efficiency, such as allowing more than one particle to occupy a terminal sub-cell (see \citet{wadsley2004} for a discussion). Note that most trees are not momentum conserving (they violate Newton's third law of equal and opposite forces); fast multipole methods (FMM) can be momentum conserving, but are more complex, and still approximate the force \citep{greengard1987}. A big advantage to using a tree for computing forces is that it can also be used to find particle neighbours in $N \log (N)$ time or better (the moments are not needed for this). Parallelising tree codes can be complex, but a basic strategy is to use a balanced spatial tree to distribute equal work among processors, then each processor has its own tree for its particles \citep{wadsley2004}.

Another approach, though following the same basic idea (details in \citet{sanchez2011,sanchez2012}), is to divide the simulation space with a static grid and keep the distances between their geometrical centres, and their powers, in memory so that look-up tables can be used.  The static-grid geometry takes advantage of the fact that the simulations of self-gravitating aggregates usually requires the particles to be concentrated in a region of space and not disperse.  Whereas the data kept in memory avoids unnecessary, repeated calculations.

\subsubsection{Monodisperse, polydisperse spheres, and other shapes}
\label{S:disperse}

In nature, grains are found in various sizes and shapes, from round beads to complex anisotropic particles. The morphology of the individual grains impacts strongly the global behaviour of the entire granular media \citep{cleary2002,pena2007,lu2007,donev2004}. Moreover, phenomena such as convection and segregation may arise in polydisperse systems \citep{kudrolli2004,dziugys2009,metzger2011,rietz2008,miyamoto2007}, see Sect.~\ref{S:segregation}. Since the handling of the contacts in SSDEM relies on sphere interaction, a common method to build particles of complex shape is to agglomerate several spheres~\citep{ferellec2010,thomas1999}. For this, spheres of different sizes are glued together allowing large overlaps in order to minimize bumpy edges along the surface. This procedure can be used to create macroscopic rugosity \citep{ludewig2012} as well as anisotropic particles like ellipsoids \citep{vuquoc2000} and even more complex bodies.

Regarding the algorithm, only few modifications have to be made in order to simulate complex shapes. Indeed, the contact detection between two particles remains a contact detection between their respective constitutive spheres. Moreover, the mass and the inertia matrix of each agglomerate are calculated at their creation so that the different forces and momenta can be easily integrated during the simulation. However, a major drawback remains: the large number of spheres that is required to model each particle will strongly impact the computational time.   Along with this, the increased complexity of the equations of motion for the aggregates and the additional memory requirements which makes parallelism very difficult.

Another approach to simulating realistic granular systems and the behaviour of non-spherical is the implementation of rolling and twisting friction \citep{ai-chen2011}.  This still relies on spherical particles, but the additional friction terms will change the way they rotate on top of one another to mimic the individual and bulk-behaviour of non-spherical particles. DEM methods were adapted to include grains of polyhedra shapes. The main difficulty when moving from a sphere-based grains to more complex shapes is the implementation of a fast contact detection algorithm. This can be done thanks to development obtained in the video game industry  \citep{movshovitz2012}.
Lastly, in the CD method, exact particle shapes can be used since the constraints of mutual exclusion and dry friction do not refer to particle shape. Thus the CD method allows for considering general polyhedral shapes, and other complex-shaped particles with multiple contacts \citep{azema2017}, but with considerably increasing computational cost. 

\subsection{Limitations}
\label{S:limitations}

As any other numerical method that has been developed in order to study a physical system, the methods presented here have limitations which have to be taken into account, so that results are not misinterpreted and methods are not misapplied.

First off, one of the main common limitations of DEM codes is the number of particles that can be simulated given the computational and time constraints.  Though some naturally occurring granular systems, and in particular laboratory experiments, can be replicated and simulated almost grain by grain, many more consist of more particles than is possible to simulate with the current computational facilities available for research.  Therefore, a compromise must be reached between the number of particles that form a real system and those that can be simulated for any practical purpose.  Ideally, we want a simulation with enough particles so that some characteristic parameters such as angle of friction, angle of repose, cohesive/tensile/compressive strength, bulk density and filling fraction---among the most commonly used---match the real system.  The inclusion of self-gravity in the calculations only exacerbates this problem. DEM methods however can handle a large number of particles with the aid of HPC computing. 

Another common limitation is the size distribution and shapes that can be used for the particles.  In general, very wide size distributions are not simulated as the number of particles required to have a representative system grows to impractical levels.  As an example, for a 50/50 mixture by volume of 1\,mm and 1\,cm spherical particles, for every 1\,cm particle particle there are up to $\approx$740 of the small 1\,mm size particles, depending on the packing.  Additionally, the contact detection algorithms rely on the search for the closest neighbours to avoid unnecessary checks and for that the simulation space is divided into cells that are large enough to contain the largest particle.  A large size disparity means that any given cell will contain a great number of small particles that are too far away to collide.  However, the contact detection algorithm will be forced to search for possible contacts when almost none is possible.  As for the shapes, an increasing complexity in the particle shape means more sophisticated contact detection algorithms and equations of motion, which will in turn require greater computing time. Besides, \citet{procopio2005} analysing the case of compaction with multiple particle FEM method, showed that compaction is not only porosity driven. The mechanical response of an assembly of particles during dense packing with finite elements models (FEM) is softer than in discrete element models (DEM), which are generally limited to non-interacting contacts, or need to be corrected to allow densities larger than approximately 0.8 \citep{harthong2009}. 

For HSDEM, the premise that the free time path is much larger than the duration of collision, $t_f\gg t_c$, is of paramount importance as this determines the systems that can accurately be simulated with this method.  More specifically, this method can be used to simulate the behaviour of molecular and granular gases, systems in a collisional regime.  The premise begins to be untrue for systems in frictional regimes (granular liquids) and condensed phases (granular solids).  In these latter regimes, the particles sustain long-lasting contacts and the approximation of $t_c=0$ is no longer valid.  Furthermore, if applied to these systems, unless the restitution equations are modified, the particles, not being able to stay in contact, will try to collide infinitely often in a finite amount of time.  This is what has been defined as {\it inelastic collapse} \citep{mcnamara1994}.  Even with this modification however, unless in a crystalline structure, the system will have an angle of friction between $0^o-5^o$ and will behave as a granular liquid.

For SSDEM one specific limitation is the time step ($\delta t$) needed for the integration of the equations of motion.  The magnitude of $\delta t$ is determined by two characteristics of the simulated system: the stiffness of the grains - given by the contact model (linear spring-dashpot, Hertzian spring-dashpot are commonly used), and the characteristic collision speed.  In general, in SSDEM codes the stiffness of the grains is underestimated in order to avoid a too small $\delta t$ as it is chosen as a fraction of $t_c$.  Using a $\delta t$ which is too large compared to $t_c$ would result in missed collisions or particles overlapping excessively as the collision dynamics cannot be accurately resolved. On the other hand, the grains cannot be so soft that the overlap of the particles is too great for a typical collision  (1\% overlap is common). It is  worth mentioning that changing the stiffness does fundamentally change the material behaviour; if it is important to the problem, this parameter can be tuned to better capture the critical physics (e.g.~sound speed), at the expense of a smaller maximum overlap and therefore a smaller timestep. Thus, a compromise must be reached so that the simulation is still realistic without being impractically slow.  This makes this method much slower than HSDEM for highly active, rarefied systems and much better suited for more densely packed, quasi-static systems.

As to the CD method, it has the advantage of allowing for much larger time steps than the SSDEM but needs sweeping the contact network a number of times in order to determine the contact forces and particle velocities at each time step, until the convergence criterion (precision on forces and velocities) is reached. The number of iterations increases with both the number of particles and the required precision \citep{Radjai2009}. Hence, in practice, CD simulations can be either much faster or slower than SSDEM simulations, depending on the precision used. Obviously, the CD method is more adequate than the SSDEM when the ratio of external stress to the particle stiffness (leading to small particle deformations) is small and when the real stiffness matters for the physical behaviour of a granular material. Moreover, except for spherical particles for which the Hertz force law is classical, there is no general force law for contacts between particles of arbitrary shape (for example, for face-face contacts between polyhedral particles). For this reason, in applications of the SSDEM arbitrary particle shapes are generally modelled as aggregates of spherical particles.  
Besides, it is also important to consider the parallelisation potential of different methods. To this respect the CD method, being based on a global determination of forces by iterations, cannot be as efficiently parallelised as the SSDEM.

%% file: AARv_NUMAGS_S5.tex
\section{Experiments}
\label{S:5}

\subsection{Lab experiments on ground}
\label{S:ground}

Ground-based impact experiments have provided fundamental insights into granular flows.  The experiments on a wide range of flows provide a basis for understanding granular flows on asteroids, but with the strong caveat that a number of observations are linked to ground-based environmental conditions very distinct from small bodies in space. In particular, humidity and the air surrounding granular matter in most ground-based experiments affect the behaviour of granular systems. For example, analyses of ejecta flow showed the importance of interstitial air in driving very high ejecta flows \citep{lohse2004}. Moreover the gravity field of the Earth is predominant, thus experiments in vacuum and in micro-gravity would be closer to reality. 

Since there are a number of possible interaction forces between grains, ground-based experiments also provide important insights into the relative importance of various physical mechanisms in granular materials.  Here, intuition from molecular or atomic systems often leads our intuition astray.  For example, collisions between dielectric molecules tend to neutralise the system for molecules or atoms of equal type.  It requires mixtures of different materials to charge molecular systems through collisions.  However, \citet{RN4095} found that granular particles of equal types, colliding under dry conditions, can accumulate charge. The key insight is that during collisions of granular matter only the areas near a contact point neutralise. Thus in the presence of electric fields (e.g. due to charges of neighbouring particles) collisions can charge particles.  

Similarly, \citet{Shinbrot8542} discovered through ground-based experiments that the features of Martian gullies typically associated with fluid immersed granular flows, may also be observed in dry flows when the flow speeds are high compared to the typical settling speeds, an effect expected to be enhanced under the reduced gravity conditions of Mars.  

For slow granular flows and plastic deformations of granular matter, where grains remain in contact with each other and rearrange through rolling or sliding, gravity in the bulk of the flow is small compared to contact forces between particles. Thus ground-based experiments will yield flows that are also expected under microgravity conditions.  New experiments allow us to measure translations of all particles in three dimensions in slow granular flows \citep{doi:10.1063/1.3674173}.  First studies on these systems have provided insights into the important question of reversibility and ageing of granular matter at the particle level.  More recent experimental studies also allow for analysis of the rotations of particles within a three dimensional granular flows \citep{Harrington2014}.  

Finally, earth-based experiments have revealed one of the most ubiquitous differences between granular matter and atomic or molecular matter: segregation of particles by size, weight, or shape. 

\subsection{Segregation}
\label{S:segregation}

An intriguing feature of polydisperse granular matter is their tendency to segregate when submitted to external constraints. 
The phenomenon is omnipresent in nature but also in diverse processes implying the handling of granular matter. From an industrial point of view, segregation is often considered as a parasitic effect that hinders the homogeneous mixing of components \citep{poux1991} especially since it can be triggered by any variation in mechanical properties of the grains. Despite most of its driving mechanisms are related to gravity, granular segregation is also observed in low gravity environment where it is held responsible for particular surface granulometry of small celestial bodies covered by regolith \citep{miyamoto2007,asphaug2009,gundlach2013}. Generally speaking, granular segregation is observed and studied in various situations. Segregation is observed in both rapid granular flows and plastic flows for almost all conditions \citep{ottino2000,jaeger_1996}, and often accompanied by convective flows \citep{rognon2010}. For mixtures of large and small, or heavy and light particles the tendency of granular flows to segregate manifests itself in the Brazil Nut Effect (BNE), where large or heavy particles rise to the top.  For plastic flows driven by periodic forcing, the onset of segregation with increasing forcing amplitude was seen to coincide with the onset of convective flow \citep{harrington2013}.

When grains are slowly poured onto a plate a heap forms and starts to grow. However, the slope angle between the pile's surface and the plate will never exceed a limit value called the angle of repose that depends on the size, shape and rugosity of the grains. When a binary mixture flows down a heap, segregation occurs since the different mechanical properties of both species lead to different angles of repose. This phenomenon is known as granular stratification and results in the formation of layers in which grains of different species are separated \citep{makse1998, koeppe1998, aranson2006, fan2012, shimokawa2015}. The formation of strata is linked to the avalanches along the heap's surface. As the granular matter flows down the heap, voids are created along the surface. These voids are more likely to be filled by small grains with creates a downward flux of the latter while large grains remain on the top \citep{savage1993, cizeau1999, kudrolli1999, gray2005,schroeter2006, gray2011}. Each of the so obtained pair of layers grows, by the propagation of a kink, from the bottom to the top of the pile.

Segregation can also be studied in a rotating cylinder partially filled by a granular mixture. Its this case, radial segregation occurs quite rapidly: small and rough particles migrate towards the centre while large and smooth grains rotate around them \citep{khakhar1997, hill2004, hajra2011}. Under certain conditions, the radial core develops more complex patterns \citep{khakhar2003, zuriguel2006}. If the cylinder is long and narrow, radial segregation is often followed by axial segregation where patterns of segregated bands appear along the axis of rotation \citep{zik1994, clement1995, cantelaube1995, caps2003, fischer2009}. This phenomenon is well known and was observed for the first time by \citet{oyama1939}. The mechanics of axial segregation are related to the different dynamic angles of repose of the rotated grains, i.e. the angle of the slope in the drum for  continuous flow regime \citep{makse1999, orpe2011, seiden2016}. Moreover, axial segregation has recently been observed in the case of a spherical container, rotating about its horizontal axis \citep{finger2016}.

When a granular mixture is vibrated vertically (i.e. on Earth's gravity), larger particles rise to the top of the system. This phenomenon, known as Brazil Nut Effect, has been studied for a long time \citep{rosato1987, mobius2001, garcimartin2002, godoy2008, metzger2011, matsumura2014} and is linked to mechanisms such as percolation and granular convection \citep{knight1993, hong2001, huerta2004}. Under the effect of vertical shaking, the particles in the system lift off. The voids that are created that way are easily filled by the small grains which leads to a ratchet-like rise of the large ones. Moreover, the vibration induced convection in the container creates a wide upward flow in the central part of the system but only a thin downward flow along its boundaries so that all sorts of grains can rise to the surface but only small particles can dive back to the bottom. Investigations during parabolic flights \citep{guettler2013} have shown that BNE can be observed in low gravity conditions even though its driving mechanisms are strongly linked to gravity, although sometimes in a complex manner \citep{staron2016}. This result consolidates the theory that certain surface structures on asteroids are created by this kind of segregation \citep{miyamoto2007}.

Segregation has also been observed in granular gases \citep{poeschel2003}. Unlike in continuous media, thermal agitation is not enough to generate the motion of the particles composing a granular gas. Furthermore, collisions between grains are dissipative so that external energy has to be injected permanently (often through vibrations) into the system in order to maintain a stationary gas like regime. Depending on the filling properties and driving mechanism of the system, granular gases exhibit intriguing phenomena such as anomalous scaling of pressure and non Gaussian velocity distribution \citep{rouyer2000, losert1999, tatsumi2009, falcon2013, scholz2017}. If one stops the external energy supply, the average energy in the system decays which is known as the cooling of a granular gas. After a while,  slow and dense regions called cluster form in the cold regions of the system \citep{goldhirsch1993, mcnamara1994, maass2008, brilliantov2018}. In the case of binary granular gases it has been shown that clustering can be followed by a particular kind of segregation where domains of the same granular species tend to merge together \citep{cattuto2004}. In denser granular gases, clustering can occur despite of an external energy injection \citep{falcon1999, noirhomme2017}. In these systems, a large and slow domain forms in the centre of the container and acts as a liquid phase coexisting with a surrounding gas phase. Here again, in the case of a binary mixture, clustering goes hand in hand with segregation. It has been shown that the different granular species segregate within the cluster, giving rise to a layered structure of the bulk \citep{serero2009, opsomer2014, opsomer2017}.

\subsection{Experiments in microgravity}
\label{S:microg}

\begin{figure}
  \includegraphics[width=16cm]{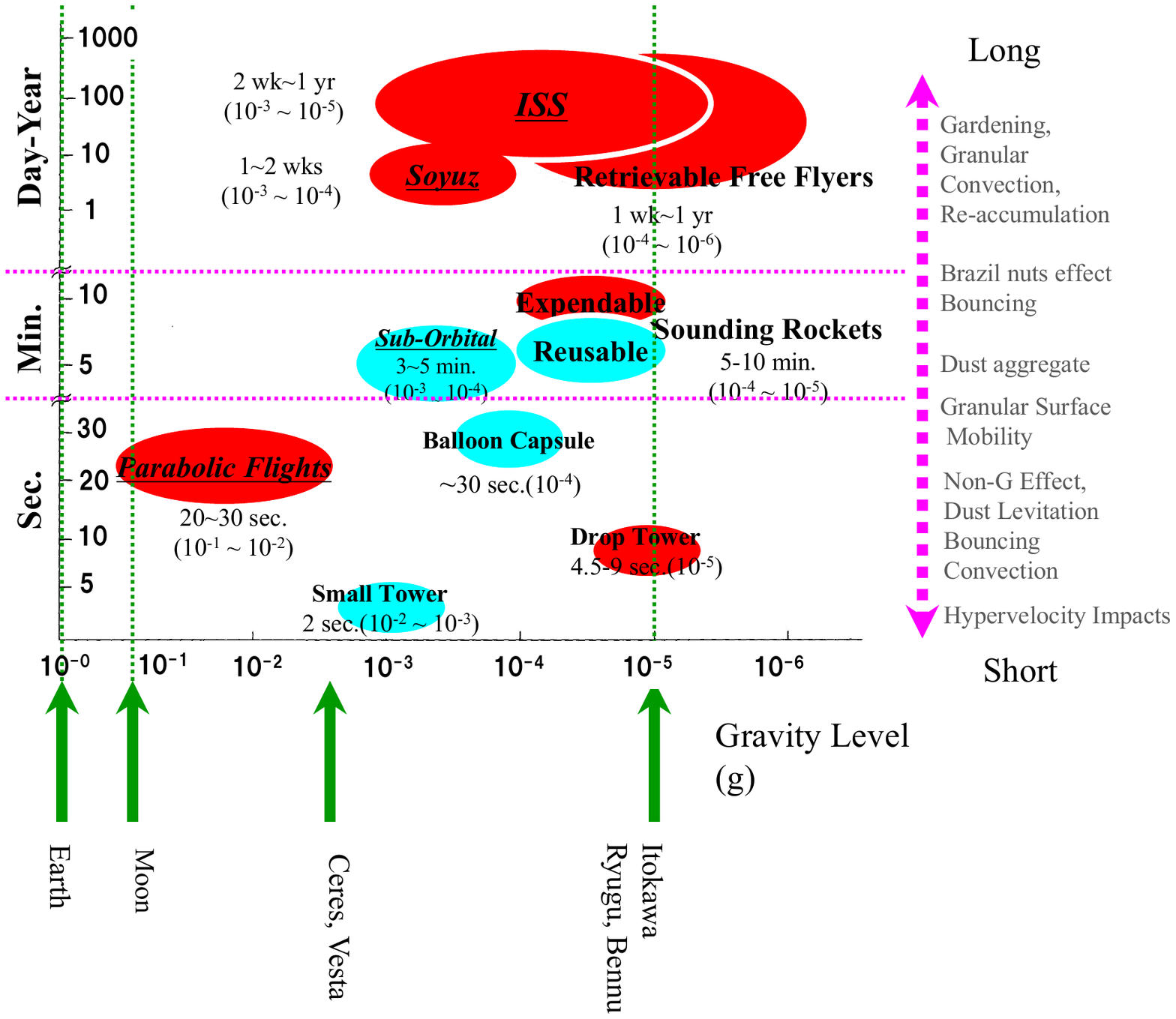}
\caption{Schematic summary of low- to micro-gravity experiments available, with different duration of the gravity environment. }
\label{fig:microgravity}       % Give a unique label
\end{figure}

Microgravity experiments can be realized by several means. However, in all of them, the key mechanism to approach weightlessness is free falling. Indeed, experiments in droptowers, in parabolic flights, and even in the international space station are falling down towards earth all together with their measuring instruments \citep{pletser2004,vonkampen2006}, see Fig.~\ref{fig:microgravity}. Granular materials have been studied in microgravity for about two decades. In particular, granular gases have been the focus of this research since the microgravity conditions allow for a more homogeneous distribution of the particles in the system \citep{poeschel2003,heisselmann2010,hou2008,leconte2006}.  In the late nineties, \citet{falcon1999} observed for the first time clustering of a continuously driven granular gas during the Mini-Texus 5 rocket experiment. Since then, the transition from granular gas to cluster has been investigated numerically \citep{opsomer2011,noirhomme2017} and experimentally \citep{maass2008,tatsumi2009}, and is nowadays in the focus of the VIP-Gran Topical Team of the European Space Agency\footnote{SpaceGrains ESA Topical Team from the European Space Agency. \tt https://spacegrains.org}. Experiments with driven rod shaped particles were realized in a REXUS flight. Thanks to a precise tracking of the anisotropic particles, non-Gaussian velocity distributions and energy non-equipartition could be highlighted~\citep{harth2013}. In another experiment~\citep{lee2015}, free falling charged particles have been studied. The authors report the observations of individual collide-and-capture events between particles, including Kepler-like orbits. Other parabolic flights have studied granular convection in varying gravitational conditions \citep{murdoch2013c}, and shear reversal in regolith dynamics \citep{murdoch2013b}. Bouncing and cohesion through Van der Walls force on aggregates or clusters of small grains ($<1\,$mm) have been tested on ZARM droptower \citep{brisset2017} at $10^{-6}\,$g, as well as on sub-orbital rocket down to $10^{-3}\,$g \citep{brisset2016}. Low velocity impacts (2-40 cm/s) of larger (approx. 10cm) projectile on a granular surface have been tested using an Atwood machine installed in a small droptower \citep{sunday2016}. This system, which uses a system of pulleys and counterweights, allows the effective surface acceleration of the granular material to be varied from $0.2-1\,$m/s$^2$ \citep{murdoch2017}. The latter experiment showed shallow penetration ($<$1/4 of the projectile diameter) and no rebound; nevertheless further experiments using other surface materials, impactor properties, and gravity regimes could be performed to confirm this behaviour.

Segregation mechanisms of denser systems with an intruder were also studied during parabolic flights \citep{guettler2013}. 
A large particle is placed at the bottom of an assembly of smaller ones. Through the shaking of the cell, the intruder rises up to the surface of the pile (the Brazil Nut Effect, see Sect.~\ref{S:segregation} above). The uprise speed was then measured for different values of gravity (Earth, Moon and Mars gravity) and a first scaling law was proposed. In bi-disperse granular media, segregation phenomena can occur even in microgravity \citep{louge2001,opsomer2017}. The driving mechanism is no longer convection and percolation as on Earth, but rather the gradients in the fluctuation energy of the grains.  Though not many experiments involving granular matter have been carried out to date in the ISS, the Strata-1 experiment \citep{fries2016,fries2018} was the first to put different mixtures of grains in orbit for a long period of time.  The samples ranged from spherical glass beads to glass shards, to crushed meteorite simulants and even a real crushed meteorite sample.  At the moment, the results of this experiment are still being analysed \citep{dove-lpsc2018} and the Strata equipment is being repurposed as the Hermes facility \citep{john-lpsc2018} that will be made available to other researchers in the near future.  One of the advantages of running experiments in the ISS is the long duration and quality of the microgravity environment which are essential in order to observe the evolution of systems that take more than a few seconds to be finalised and therefore, could not be carried out in droptower or parabolic flights due to their short duration.

%% \todo{Hajime: maybe we will need a higher resolution for this figure}

%% file: AARv_NUMAGS_S6.tex
\section{Benchmarking cases}
\label{S:6}

\subsection{Macro- and micro-scale benchmarks: what to learn from observations}
\label{S:scale}

Theoretical development, numerical modelling, and experiments of granular systems in low- and self-gravity regime, all concur to a better understanding of asteroids. This is often needed for studying sample return mechanisms, for estimating the outcomes of high-velocity impacts, collisions and ejecta, or for predicting low-velocity impacts and bouncing of platforms on the surface, etc. \cite[e.g.][]{biele2017, thuillet2017, murdoch2017, ballouz2017}. In this section, we will focus in the following on segregation phenomenon, long-term evolution spin-up and mass shedding, and analysis of the reaccumulation process. Several computer simulations have been proposed to tackle the analysis of self-gravitating planetary bodies as granular systems, starting mostly with HSDEM \citep{richardson1993,richardson2011,leinhardt2000,murdoch2012,michel2002,walsh2006,tanga2009b,walsh2012,bagatin2018}, and later with SSDEM \citep{sanchez2011,schwartz2012,tancredi2012}, see Sect.~\ref{S:4}.

\subsection{Segregation}
\label{S:segregation-bench}

Shaking that gives rise to segregation (see Sect.~\ref{S:segregation}) can have different origins on asteroids (reaccumulation, impacts, tides, stress cycles, ...), with different frequencies and amplitudes. Seismic shaking has been proposed as a mechanism to resurface asteroids and account for the presence of ponds on (433)~Eros \citep{cheng2002,richardson2004}, or boulders on (25143) Itokawa \citep{saito2006,miyamoto2007}, or young surfaces on specific classes of Near-Earth asteroids \citep{binzel2010}. In addition to flows, the BNE could then be effective on asteroids and provide segregation on regolith grains. This has been simulated numerically by methods presented in Sect.~\ref{S:4} \citep{sanchez2009,tancredi2012, murdoch2012, matsumura2014, perera2016, maurel2017}, with hard-sphere DEM, and later soft-sphere DEM. \citet{perera2016} suggest that the mechanism of the BNE that is most relevant---in presence of a  binary size-distribution---is that of percolation, and not convection, and that the innermost regions remain unsorted. Further, \citet{maurel2017} have performed numerical simulations investigating the BNE in an unconfined environment. They show that, under Earth gravity ($1\,g$), a void-filling mechanism is predominant, in contrast to more classical granular convection-driven BNE in the presence of walls. While this void-filling mechanism remains relevant in a lower gravity regime ($10^{-4}\,g$), it is however differently influenced by the friction properties of the particles. Last, \citet{chujo2017} have analysed the BNE (as well as the reverse BNE, occurring when oscillation frequency is high, and the bulk density of the larger particles is larger than that of the smaller ones) in a low-gravity environment with an intruder, under `less-convective' conditions. They also point out that the amplitude and frequency of vibrations that may be induced on small bodies is still not well known.

\subsection{Rotation, spin-up, mass shedding}
\label{S:spinUp}
 
\subsubsection{The YORP Effect:}
It has been generally accepted that solar radiation pressure, the absorption, re-emission and reflection of photons emitted by the Sun, can, over the lifetime of SSSBs, produce not only an acceleration of its orbit (the Yarkovsky effect) but also a net that will change their rotation state \citep{rubincam2000}.  For this torque to appear however, it is essential for solar photons to impact a body with an irregular shape as otherwise photons would be re-emitted or reflected in symmetric directions and their effect would average out.  An asymmetric body on the other hand would be similar to a propeller; sunlight bouncing off the blades and causing it to change its rotation.  This is what is defined as the Yarkovsky-O'Keefe-Radzievskii-Paddack effect (YORP for short).

\subsubsection{Deformation and Disruption}
Given that SSSBs are granular in nature, their structure can yield/fail under stress.  Events that can cause this structural failure could be rotation at high enough spin rates $\Omega$, collisions and planetary flybys.  From these, in this section we will focus our attention on the failure that is produced due to high rotation rates.

As previously explained, the interaction of an irregularly shaped body and solar photons, over millennia, can produce a net torque that can change the rotation rate of a SSSB.  Though it is possible to calculate the critical rotation period for a SSSB of arbitrary shape numerically, it is useful to start with a simpler shape so that we can obtain some insight.  For this, we could start with a sphere of radius {\it r} and bulk density $\rho$.  

For a gravitationally bound aggregate, the critical spin period \[P_c = 2\pi/\Omega_c\] can be calculated by equating the acceleration of gravity at the surface of a sphere with the centrifugal acceleration at the equator, just as it was done by \citet{pravec2000}.  This would allow us to derive a criterion for the critical limit of rotation period, depending only on the density of the sphere $\rho$,
\begin{equation}
\frac{Gm}{r^2}=\Omega_c^2\,r\rightarrow P_c=\sqrt[]{\frac{3\pi}{G\rho}}
\end{equation}
where $G$ is the gravitational constant and $m$ is the mass of the sphere.  For a bulk density between 2000--3000\,kg.m$^{-3}$, this expression gives us a spin period $P_c$ between $2-2.3\,$h.  Conversely, the spin barrier of the observed asteroid population is $\approx 2.4$ h (see Sect.~\ref{S:observations}, Fig. \ref{fig:lcdb}). Though this simple calculation provides a first approximation into the make-up of asteroids, there is a feature that it can not explain.  For the smaller asteroids (below $\approx 100-200\,$m), this spin barrier does not seem to exist \citep{pravec2000}.  Additionally, this spin barrier might more accurately be described as a deformation or a fission barrier, as asteroids reaching it would have to deform or go through a fission process as their structure would fail at such high spin rates \citep{jewitt2013,jewitt2014,hirabayashi2014c,walsh2008,sanchez2016,hirabayashi2014d}.

From here, we can take one step further and study what happens to tri-axial ellipsoids when subjected to high spin rates.  The form in which these bodies would fail can also be solved analytically, though this requires the use of Soil Mechanics theories---elastic-plastic theories to be more precise \citep{holsapple2001,sharma2009}---in particular, limit analyses.  Within this approach, one seeks the maximum load that a body (whatever its nature) can sustain without failing.  This circumvents the need to know the past history of the body.  For these theories to be applied, a framework is needed; a basic principle for many granular (soil or gravel) or solid (rock) geological materials is that the shear yield stress on any plane increases with the normal pressure on that plane. The simplest criterion of that type is the Mohr-Coloumb (Drucker-Prager is another, more sophisticated possibility) yield criterion \citep{holsapple2001}, determined solely by a cohesive strength and an angle of friction. Also required is some `flow rule' to determine the flow: some prescription of the plastic deformations that occur when the stresses meet or exceed the yield criteria. Then, in principle, it should be possible to trace a particular loading history and determine the resulting stress fields.

By doing this, it was possible to determine that most asteroids shapes and rotation rates were well within the equilibrium limits for low angles of friction which are typical for dry, cohesionless soils and so the observed shape and spin rate of the asteroid population could not be used to assert that these bodies had to be cohesive.  This analysis however, could not be applied to comets where small cohesive forces could provide additional strength to the structure.

This, of course would mean that the source of the strength of asteroids is purely gravitational -- in the case they are not monoliths -- and that all observed asteroids should be below the spin barrier.  However, it was observed that the small asteroid population\footnote{Sizes, when not measured directly, are estimated from the absolute magnitude $H$ and by assuming an albedo of 0.2} (approximately $< 100-200\,$m) contained not only fast, but even super-fast rotators with periods of only a few minutes possibly depending on their taxonomic type \citep{pravec2000,warner2009,taylor2012,perna2016}. This is what has led scientists to believe that cohesive forces could be important for these granular systems; or, otherwise, that the small fast-spinning asteroids are essentially monolithic rocks \citep[e.g.,][]{polishook2016}.

At this moment we have to make a point about nomenclature. We will use the word cohesion and cohesive force to refer to the surface-surface attractive force between any two grains in contact.  The term cohesive strength will be used to refer to the parameter in the different yield criteria which is defined as {\it the shear stress at zero normal stress}.  Complementary to this, the term tensile strength is defined as {\it the normal stress at zero shear stress}.  These two could be related for specific materials, but there are indications that their relationship can change at very small loads \citep{kim2009}.

The deformation of ideal spherical and ellipsoidal self-gravitating aggregates can also be studied with the same theoretical tools presented above \citep{holsapple2001,holsapple2004} and extended to allow for cohesive and therefore, tensile strength \citep{holsapple2007}.

By doing this, it was found that the presence of tensile and cohesive strength for a large body ($>10\,$km) makes no difference in the permissible spin.  That is, gravitational effects dominate the strength of the body.  This implies that the observed spin limit for large bodies cannot be used to infer zero-strength (cohesive/tensile) granular asteroids. On the other hand, the strength that allows the higher spins of the smaller and fast-spinning km-sized bodies is only on the order of $10-100\,$kPa, a very small value compared to small terrestrial rocks.  Additionally, it is stated that the strength needed for small granular asteroids to become fast rotators could be originated in an accumulated slight bonding between their constitutive particles.

For asteroids between approximately 0.2 and 10\,km, cohesive strength does not seem to influence the maximum spin rate they can reach, but it affects the overall shape of the bodies.  This could partially explain the shapes and surface morphology of asteroids Itokawa, Ryugu and Bennu for which photographic evidence is available.

Though the theoretical results described above provided great insight into the mechanisms that shaped the observed asteroid population as well as their internal structure, questions about the origin of their cohesive strength as well as the origin of binary asteroids, and asteroid pairs were still unanswered.  Other questions, more related to planetary defence and space exploration, also needed answers. 

One weakness of the theoretical models used in the above described research is that they are not dynamical. That is, they cannot show the reshaping of the bodies and are always constrained to ellipsoidal shapes.  It is here where the use of DEM codes becomes important.  The use of these kinds of codes followed the same path as the theoretical efforts, that is, start with spherical and ellipsoidal shapes, no cohesive strength and the added numerical particularity of having only mono-disperse spherical particles as the constituents of the aggregates.  A drawback of this last point was that crystallization was unavoidable, but even this was used to emulate higher angles of friction \citep{walsh2008} when needed. These aggregates would reproduce the behaviour of ensembles of particles with friction angles of $\approx 40^o$ when crystallized and between $0^o-5^o$ when the particles were randomly packed (not naturally found for granular matter).

The influential work of Walsh et al \citep{walsh2008,walsh2012} provided evidence supporting the idea that binary asteroids could be continuously formed through the shedding of asteroid material at high enough spin rates and its subsequent reaccumulation to form the secondary of a binary system.  However, there were questions about the needed time for this process \citep{jacobson2011} to take place as well as the influence of the crystallization of the aggregate. Besides, the final axisymmetric top-shapes may be a particular outcome from the simulation of \citet{walsh2008}, as shown by \citet{cotto2015} who analysed the coupled spin-shape evolution with different initial configurations of the aggregated particles, and reducing crystallisation effects. Including polydisperse spheres and different block shapes adds extra complexity and realism to the simulation models \citep{walsh2008,walsh2012,michel2013,bagatin2018b}.

Subsequent studies carried out by \citet{sanchez2011} using a SSDEM  code  did implement random packings---as in \citet{comito2011_epsc}, and in contrast to \citet{walsh2008}---so that crystallisation was explicitly avoided, and they started to study the influence of surface-surface friction. Their studies showed that this type of simulation fully agreed with the theoretical models of Soil Mechanics \citep{sanchez2012} for aggregates with angles of friction of $\approx 12^o$ and $\approx 25^o$ (these angles of friction are not naturally found in nature for gravel).  That is, the aggregates followed the deformation path that was determined by the theory and at the correct spin rates.  Additionally, it was found that the aggregates failed at the centre, not producing granular flows on the surface which was the failure mode showed by \citet{walsh2008}.  One intriguing feature was however, that at times the aggregates would split into two almost symmetrical pieces and some others, it would simply eject individual particles.  Besides, \citet{tanga2013} showed that particle ejection is not incompatible with splitting, as the two can occur in the same system if the spin-up phase is continued. 

\subsubsection{Cohesive Strength} 
Even though it had already been established that a hypothetical fast rotating gravitational aggregate would necessarily have some cohesive strength, its source had not been established.  To tackle this problem, the work of \citet{scheeres2010} attributed this to the comparatively strong van der Waals forces.  If we define a {\it bond number {\bf B}} to be the ratio between the cohesive/adhesive attachment a particle feels and its weight in an asteroid environment, this ratio becomes 1 for centimetre-size particles in the ideal case. Of course small SSSBs with high spin rates can either be strong monoliths or gravitational aggregates with relatively weak cohesion; indeed, the observed spin rates of fast rotators do not require the high cohesive strength of solid rock \citep{sanchez2014}.

Supported in these findings, \citet{sanchez2014}, using the same DEM code, find that the tensile strength of an ensemble of self-gravitating spherical particles is inversely proportional to the average particle size.  This result allows them to study self-gravitating aggregates with realistic asteroid sizes, friction angles and tensile strength, but without the burden of having to simulate an impractically large number of particles \citep{sanchez2014}. Angles of friction of $\approx 35^o$ were obtained through the implementation of rolling resistance \citep{ai-chen2011}. In essence, in their model, the larger boulders are embedded in a cohesive matrix that holds the entire aggregate together.  In a way, this would be a van der Waals cement.  They calculate that a cohesive strength between 25-100 Pa would be enough to explain why asteroids approximately $<100-200\,$m in size could have spin periods below the 2.2 h of the spin barrier. Other works compared the hard-sphere and soft-sphere DEM simulations including cohesion as well \citep{schwartz2013}.  Studies of observed fast-rotating asteroids also confirm to have cohesive strengths that are far below what would be expected for competent rocks \citep{rozitis2014,hirabayashi2014d,hirabayashi2014c}. 

Additionally, a subsequent study by \citet{sanchez2016} found that the amount of deformation of a self-gravitating aggregate is greater for low angles of friction and vice-versa.  Whereas the amount of cohesion was directly related to the fission process.  That is, at low or no cohesion, particles would be ejected in a one-by-one-fashion, producing a tail.  Whilst at high cohesion, the aggregates would eject larger, coherent groups of particles at once.  This could result in the aggregate going through a catastrophic disruption process that would break the body in several coherent pieces.  For the most cohesive cases, the aggregates would split in two symmetrical pieces.  Similar work by \citet{zhang2018}, incorporating friction and cohesive interactions in the manner suggested by \citet{sanchez2016}, confirmed most of these results.

\subsubsection{Internal Structure}
Up to this point, one assumption made by most studies is that the modelled aggregates and asteroids had completely homogeneous interiors.  However, this is not necessarily the case. For instance, it is also believed that in a post-catastrophic collision large blocks with lower velocities reaccumulate first, and are at the centre of the body, while smaller material could be kept closer to the surface by friction \citep{britt2001}.  In this regard, it has been shown through theory and simulations that a spherical aggregate with a central core, which is structurally stronger than its external shell, would avoid the initial internal failure prevalent in homogeneous aggregates and fail at the surface \citep{walsh2012,scheeres2015,hirabayashi2014b,hirabayashi2015}.  In numerical simulations, such  structure was the only one able to reproduce the surface slope of asteroid 1999~KW4 \citep{sanchez2015} when the radius of the core was $\approx 0.7 R_b$, where $R_b$ is the radius of the aggregate, in agreement with \citet{walsh2012}, and as predicted by the theory. Those researchers also explain the surface shedding process found by \citet{walsh2008}.  Essentially, the crystalline packing used in the Walsh et al. numerical simulations had intrinsically built an interior which was stronger than the outermost shell formed by surface particles.  These aggregates were originally built to fail at the surface though the crystalline packing was used only as a tool to avoid the fluid-like behaviour of the grains when simulated with an HSDEM code. 

On the other hand, aggregates with weak cores and strong shells have been studied only through numerical simulations.  Upon rotation, these aggregates locate the greatest stress at their centre where a severely weakened interior would fail long before the stronger shell.  This means that an equatorial ridge would not be formed by granular flow ---which would not happen under these conditions--- but rather by the global deformation of the body. Preliminary results also show that if the core is about $\approx 0.5 R_b$, it is possible to obtain a shape similar to that of asteroid Itokawa, though its reproducibility has still to be proven \citep{sanchez2018}.

A possible confirmation of this correlation between fission size,  cohesion and internal structure came from a study of the equatorial cavities in asteroids 2008~EV5 and 2000~DP107 that had previously been attributed to impact events \citep{busch2011}. \citet{tardivel2018} showed that if a hypothetical original body with a filled cavity rotated at high enough spin rate, the first place to fail in tension would be exactly where the cavity appears at the moment.  Kinetic sieving \citep{gray2005}, produced as a result of surface flow, which should in turn be produced by a strong core, could produce a `rocky equator' (size segregation with a flow of the largest rocks to the equator) and explain the low tensile strength of this specific region for this fission mechanism to work.

Therefore, all these studies suggest that the shape of granular asteroids, and the specific form towards which they evolve under rotation, are intrinsically linked to internal structure and structural strength \citep{sanchez2018}.  Examples of this statement have already been mentioned, namely asteroids Itokawa, 1999~KW4, 2008~EV5 and 2000~DP107.  Additionally, late in 2013 two more observed asteroids, P/2013 R3 and P/2013 P5  were termed {\it active asteroids}.  Both structures failed, but in very different ways \citep{jewitt2013,jewitt2014}.  The former broke apart in several coherent pieces, whereas the latter exhibited several long tails.  If we assume that the aggregate and core approach can be applied to these two bodies, asteroid P/2013 R3 might have had a homogeneous interior and asteroid P/2013 P5 a strong core \citep{hirabayashi2015, sanchez2016}.

\subsubsection{Scaling} As has been proven by the authors cited above, the cohesive strength of a self-gravitating aggregate could become important for small (approximately $\le 100\,$m) bodies.  This would imply that the onset of deformation or disruption is the result of an interplay between the effects of material properties (cohesion, density, porosity, particle size distribution, friction) and gravitational forces which depend on the aggregate size and mass.  Intuitively, increasing the size of a cohesive aggregate should be equivalent (structurally) to reducing the value of cohesive strength and vice-versa \citep{sanchez-lpsc2015}.  However, the scaling was not very clear.  To solve this, \citet{azema2018} analysed the stress-strain behaviour and micro-structure of a granular asteroid, modelled as a cohesive granular agglomerate of spherical particles, subjected to vertical compression.  Based on this, they defined a modified inertial number that relates the particle-particle tensile strength (as defined by \citet{sanchez2014}) and the overburden pressure generated inside the aggregate by gravity alone. This newly defined scaling still needs to be tested against previous results.

\subsection{Post-impact reaccumulation}
\label{S:impact}

The evolution of fragments resulting from catastrophic disruption determines the distribution of aggregates that will reaccumulate by self-gravity. SPH plus N-body simulations indicate that the ejecta cloud shall typically collapse into multiple gravitational aggregates, while particles with higher velocities can escape from the system. Fragments with kinetic energy smaller than gravitational binding energy shall wind up in the largest remnant---which by definition has less than 50\% of the original progenitor's mass in a catastrophic collision. The mass distribution and angular momenta of the resulting aggregates, and how they relate to the parent body's mass and spin, are a heritage of the impact conditions under which the parent object was disrupted. Very energetic collisions, relative to the parent body's binding energy,  will result in a final aggregate of much lower mass, and composed of the fragments that were ejected at the lowest speeds. These are likely the largest fragments, though it still needs some further investigation. An off-centre collision will show its signature in a high-angular-momentum aggregate body. In any case, fragments of diverse shape and size will comprise the aggregated bodies, as shown in laboratory experiments of catastrophic collisions \citep[e.g.,][]{durda2015}. 

Numerical simulations oriented to post-catastrophic gravitational reaccumulation were initially focused on deriving the mass and size distribution of asteroid families \citep{michel2001, michel2015_ast4}. They showed that larger family members are likely reaccumulation products rather than discrete competent fragments. Gravitational aggregates obtained as a product of less catastrophic---shattering collisions---will have voids in between fragments, largely contributing to its global (macro-)porosity. An aggregate porosity is the result of the particular packing of a polydisperse collection of its components. Recently, due to the influence of granular dynamics, granular packing configurations have been investigated both experimentally and numerically---including studies in different gravitational environments. Attempts to understand the observed global shapes and structures of aggregates through DEM numerical simulation have been performed by several authors \citep{tanga2009,comito2012,michel2013,schwartz2018,bagatin2018b}. Originally, \citet{farinella1981} suggested that such a reaccumulation process would produce elongated tri-axial asteroids, following equilibrium figures of incompressible fluids. This is however not strictly the case for bodies smaller than $\sim 100-200\,$km in diameter, where shape can significantly depart from a fluid-equilibrium figure, thanks to friction and sustained shear stresses\citep{holsapple2004,holsapple2007}. However, \citet{tanga2009b} showed that external mechanisms (e.g., low-energy impacts) can gradually reshape the bulk of the body, pushing it toward a minimum-energy state while remaining compatible with observed shapes and spins. Nevertheless, a careful examination of the equilibrium figures show that these are very loosely defined by flat minima in the energy potential of the self-gravitating, rotating body \citep{tanga2009}. As such, tri-axial ellipsoids significantly far from equilibrium are possible with minimal strength.
Moreover, \citet{richardson2009} analysed resulting shapes from re-accumulation and rotational disruption, including variable material strength/cohesion and irregular pieces (modelled as `bonded aggregates') in the DEM {\tt pkdgrav} code \citep{stadel2001,richardson2000,schwartz2012}. Other simulations have been performed to describe the outcome of a catastrophic disruption on a gravitational aggregate. 
\citet{michel2013} show that the general shape of an asteroid like Itokawa, together with the presence of boulders on its surface, can be the natural result of the reaccumulation process, given some specific material parameters for the aggregate (strength, bouncing coefficient, ...).
\citet{ballouz2015} showed the influence on the catastrophic disruption threshold when taking into account the initial spin of the parent body, and some frictional effect and shear strength; but they didn't consider material fragmentation, heating and compaction from hyper-velocity impacts. 
Recent numerical simulations of the reaccumulation process by \citet{bagatin2018,bagatin2018b} were based on such {\tt pkdgrav} code. Using irregular rigid fragments instead of spherical individual particles, they study the packing fraction and the shapes of the final aggregates. They find that the bulk macro-porosity of aggregate asteroids can be related to the mass fraction of the largest component of the aggregate structure \citep{bagatin2018}. This in turn may be related to the specific energy of the collision that formed the aggregate itself. Then, some relationship between the macro-porosity of an asteroid and the kind of event that produced the object itself may be assessed. Studying the shapes of the aggregates formed at the end of the reaccumulation process, \citet{bagatin2018} find a general mechanism to explain some asteroid shapes, in particular bilobated (`contact binary') asteroids.

The reaccumulation process, while relatively short on astronomical times scales (from a few hours to several days, depending on total mass and collision boundary conditions), can be complex and result in phenomena like shaking and segregation as seen above, re-arrangements of grains and blocks and compaction \citep{ben1998,richard2005,yu2017} that can be difficult to model given uncertainties in the initial conditions of the aggregate cloud. However, numerical simulations have shown at least that bilobated shapes and satellites can be formed from continuous mass shedding \citep{walsh2008,walsh2012}, from fission, or---over a shorter time-scale---from the reaccumulation process \citep{tanga2013,schwartz2018,bagatin2018b}.

%% file: AARv_NUMAGS_S7.tex
\section{Discussions and prospective}
\label{S:7}

%\subsection{What to learn from on-going and new space missions}
%\label{S:missions}

From a scientific point of view, surveys and space missions will allow us to address many of the topics and questions raised here. As an example LSST and Gaia will bring deeper knowledge and insights on a much larger number of solar system objects. The Gaia mission, now that the second catalogue has been released \citep{spoto2018}, is promising to gather valuable data of physical and dynamical properties for hundreds of thousands of small bodies \citep{hestroffer2010_lnp}; and LSST will extend this to fainter objects, by orders of magnitude \citep{jones2009,schwamb2018}.
Possibly the first thing we can learn from on-going and new space missions is that their successes highly depend on our ability to understand their targets---small asteroids and comets---as self-gravitating granular systems. JAXA's Hayabusa and Hayabusa-2 missions, as well as NASA's OSIRIS-REx, Rosetta and DART missions have already had to deal with the complexities of these systems due to their design.  The Hayabusa mission had as one of its objectives to shoot a small pellet to the surface of the asteroid Itokawa and capture the ejecta material to bring it back to Earth as a sample. Unfortunately the system failed and the sample canister was sealed without the pellet being fired.  At its return to Earth it was found that in spite of that, some of the dust of the surface had been luckily collected.  The Hayabusa-2 mission, currently at asteroid Ryugu, did land the MASCOT and Minerva pods on the surface of the asteroid.  This manoeuvre has encompassed a very detailed study of the interaction of the landing pod with the surface of the asteroid of which almost nothing was known.  The usual characterising parameters such as porosity, angle of friction, strength or even surface density, boulder abundance, or topography are not known and the science team has had to work with a large range of parameters to make sure of their success.  Something similar can be said about the OSIRIS-REx mission, whose main science objective is to return a sample of asteroid Bennu to Earth for subsequent analysis. The Touch-and-Go-Sample-Acquisition-Mechanism (TAGSAM) had to go through a thorough design and testing process that involved not only experimentation, but also two teams carrying out numerical simulations to make sure that, regardless of the variations in the characteristics of the surface of the asteroid, the sample would be acquired and sealed safely. Of these missions, the Double Asteroid Redirection Test \citep[DART,][]{ cheng_2016} mission is the only one that has not been launched yet, and much of the work of the different working groups that form the science team has been focused on the geophysical characteristics of Didymos. Hopefully, the complementary Hera/AIDA mission will be able to complete the picture \citep{michel2018}. All these missions will give more detailed information on the surface, not only from remote imaging, but also from a gentle touch down and physical interactions  (Hayabusa, Hayabusa-2, OSIRIS-REx). They will also show how the surface and body react to high-speed impacts, cratering, ejecta generating, and transfer of linear momentum (DART, Hayabusa-2). We still need to learn more for future applications, on the possible anchoring processes, bouncing at the surface, and how the regolith can react to different tools for exploration or excavation, depending for instance on its compaction or particle size distribution. 

Unfortunately, the search for knowledge is not the only reason to obtain a better understanding about asteroid Geophysics.  A precise understanding of the composition and mechanics of asteroids is essential to guarantee that humankind can protect itself from potential impactors and other hazardous asteroids. Whether an asteroid produces a blast in the atmosphere or an impact on the ground depends---for a given size or mass---on the asteroid's global properties. Deflection demonstration missions can help resolve some of the challenges encountered in asteroid deflection by providing ground truths. The DART mission, for instance, is a NASA mission concept intended to demonstrate the change of the state of motion of an asteroid through a kinetic impact. Targeting the moonlet of the near-Earth asteroid (65803) Didymos, DART would alter the orbital period of the binary asteroid system. Measurements of this alteration would then allow us to draw conclusions on the composition and dynamics of the binary asteroid system on the one hand and kinetic impactor-based deflection techniques, on the other hand. The European space agency (ESA) is currently investigating a concept for an observer spacecraft, Hera, that would allow for a more detailed in situ assessment of the effects of the DART impact. Another deflection demonstration mission concept named NEOTwIST has been developed in the framework of EU's NEOShield project \citep{harris2013neoshield, drube_2016}. NEOTwIST would use a kinetic impactor to spin up the well characterises asteroid (25143) Itokawa \citep{eggl_2016}. The resulting change in the spin state, would provide clues as to the magnitude and direction of the momentum enhancement vector. The data acquired during deflection tests would be invaluable to gauge simulations and structure models that will be used to predict the outcomes of future deflection mission.   

Regardless of the final application of acquired knowledge, furthering our understanding of the interior, internal structure and evolution of asteroids is needed. This understanding  can be obtained through the determination of their gravity fields, tomography, and through seismic experiments and such experiments can only be obtained from future space missions.  On the other hand, laboratory experiments need to approach the conditions of low gravity, vacuum, without specific confinement, particle-size dispersion, etc., in order to better understand and model the many phenomena at play. Segregation, clustering, and internal cohesion are some of the mechanisms and properties that need further study. Numerical simulations have reached a high level of fidelity to model Earth-based experiments and can predict phenomena on low-gravity surfaces as well as on self-gravitating bodies, but would greatly benefit from experimental validation.

Having said all this, there are definitely a few things that we have been able to learn from the successes and failures in these missions: first off, that small asteroids are not only monoliths with bare surfaces.  If we start with that, other things have come to light as a consequence.  As the gravitational fields of small asteroids are very weak, cohesive, adhesive and electrostatic forces can be as, if not more, important than particle weight.  This could facilitate the formation of very porous interiors.  Since asteroids can sustain shear stresses, their shapes are not going to be regular.  Due to their formation and evolution processes, they are likely to be formed by particles with sizes that range from microns to tens of meters.  They are indeed affected by the YORP effect and, so, apart from the collisional evolution, we need to take into account their rotational evolution, strength, internal heterogeneities and individual shapes and surface features if we really want to understand them.

We have seen that our objects of study span a large space of physical and dynamical parameters, with orders of magnitudes in size and mass range, with orders of magnitude in spin-rate range, different taxonomic class which could mean different constituent materials, large differences in their bulk densities, different distances from the Sun and different thermal environments. Each asteroid is indeed a small world that can provide much insight about asteroids as granular systems, but not as much as to paint a complete picture.  Space missions and observations of specific targets remain circumstantial, which means that theoretical, experimental and modelling efforts are needed to bridge the knowledge gap.

%So, each target will teach us a lot about asteroids as granular systems, but not in a way to fully provide a complete picture. Space missions and observations of specific targets remain circumstantial, so that theoretical, experimental and modelling efforts are needed to bridge the knowledge gap.

All these examples show the need for a great interdisciplinary effort. Really, it is not only Granular Dynamics, Soil Mechanics or Aerospace Engineering, but all of these disciplines together that have some of the tools necessary to study Small Solar System Bodies, and only a true interdisciplinary effort will bring further understanding.

%\subsection{What to learn from experiments}
%\label{S:experiments}
%(including future development in experiments, e.g. vacuum and microgravity)